\documentclass[acmlarge]{acmart}

\usepackage{lipsum}
\usepackage{enumitem}
\usepackage{subfigure}
\usepackage{caption}
\usepackage{subcaption}
\usepackage{array}
\usepackage{algorithm}
\usepackage{colortbl}
\usepackage{algpseudocode}
\usepackage{makecell}
\usepackage{pifont}
\usepackage{xcolor}
\usepackage{mdframed}
\newcommand{\cmark}{\textcolor{green!60!black}{\ding{51}}}  
\newcommand{\xmark}{\textcolor{red}{\ding{55}}}             

\newcommand{\sysname}{\textit{VergeIO}}

\AtBeginDocument{%
  }

\newcommand{\orange}[1]{\textcolor{black}{#1}}

\usepackage[normalem]{ulem}
\newcommand{\add}[1]{\textcolor{black}{#1}}

\newcommand{\modify}[1]{\textcolor{black}{#1}}

\newcommand{\squishlist}
{\begin{itemize}[itemsep=1pt,parsep=2pt,topsep=3pt,partopsep=0pt,leftmargin=0em, itemindent=1em,labelwidth=1em,labelsep=0.5em]}
\newcommand{\squishend}{\end{itemize}}
\newcommand{\squishenum}{\begin{enumerate}[itemsep=1pt,parsep=2pt,topsep=3pt,partopsep=0pt,leftmargin=0em, itemindent=1.5em,labelwidth=1em,labelsep=0.5em]}
\newcommand{\squishsubenum}{\begin{enumerate}[itemsep=1pt,parsep=2pt,topsep=0pt,partopsep=0pt,leftmargin=0em,listparindent=1.5em,labelwidth=1em,labelsep=0.5em]}
\newcommand{\squishenumend}{\end{enumerate}}

\author{Xiyuxing Zhang}
\authornote{Co-primary authors}
\affiliation{
  \institution{Carnegie Mellon University}
  \country{USA}
}
\affiliation{
  \institution{BNRist, Tsinghua University}
  \country{China}
}
\email{zxyx22@mails.tsinghua.edu.cn}

\author{Duc Vu}
\authornotemark[1]
\affiliation{
  \institution{Carnegie Mellon University}
  \country{USA}
}
\affiliation{
  \institution{Michigan State University}
  \country{USA}
}
\email{ducv@andrew.cmu.edu}

\author{Chengyi Shen}
\affiliation{
  \institution{Carnegie Mellon University}
  \country{USA}
}
\affiliation{
  \institution{Zhejiang University}
  \country{China}
}
\email{chengyis@andrew.cmu.edu}

\author{Zhikai Qin}
\affiliation{
  \institution{Carnegie Mellon University}
  \country{USA}
}
\email{qinzk23@mails.tsinghua.edu.cn}

\author{Ruiqi Hong}
\affiliation{
  \institution{Wuhan University}
  \country{China}
}
\email{rachel@whu.edu.cn}

\author{Yuntao Wang}
\authornote{For correspondence, please contact Justin Chan and Yuntao Wang}
\email{yuntaowang@tsinghua.edu.cn}
\author{Yuanchun Shi}
\email{shiyc@tsinghua.edu.cn}
\affiliation{
  \institution{BNRist, Tsinghua University}
  \country{China}
}
\affiliation{
  \institution{Qinghai University}
  \country{China}
}

\author{Justin Chan}
\authornotemark[2]
\affiliation{
  \institution{Carnegie Mellon University}
  \country{USA}
}
\email{justinchan@cmu.edu}

\begin{document}

\setcopyright{cc}
\setcctype{by}
\acmJournal{IMWUT}
\acmYear{2026} \acmVolume{10} \acmNumber{3} \acmArticle{192}
\acmMonth{9} \acmDOI{10.1145/3831627}

\title{{\sysname}: Depth-Aware Eye Interaction on Glasses}



\newcommand{\reviewcomment}[1]{\vspace{0.0cm}\begin{mdframed}[backgroundcolor=gray!20]#1\end{mdframed}\vspace{0.4cm}}

\begin{teaserfigure}
    \centering
    \includegraphics[width=0.85\linewidth]{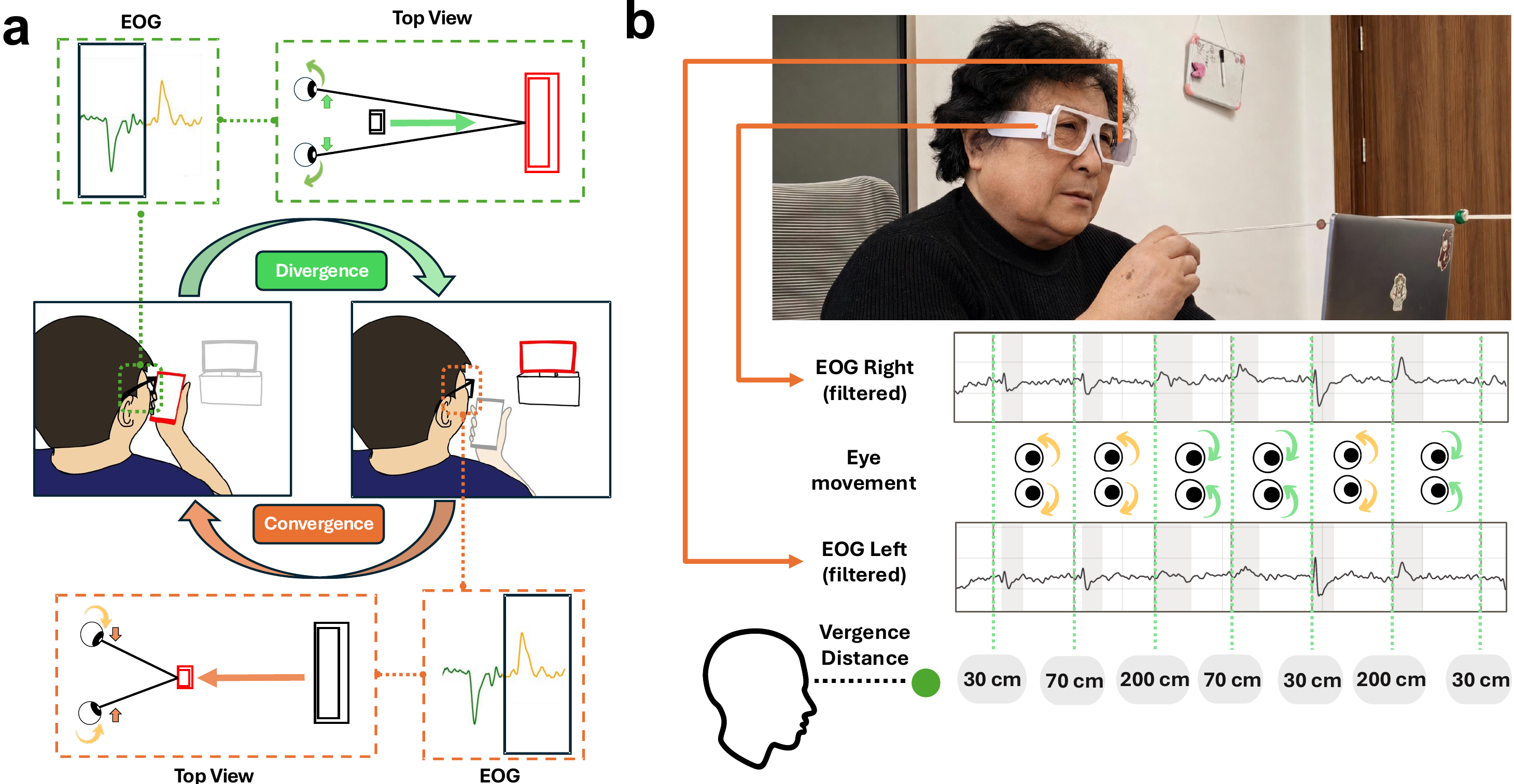}
    \caption{{\bf Overview of the {\sysname} system. (a)} {\sysname} senses eye vergence on glasses using electrooculography (EOG) by detecting gaze shifts between objects at different depths. The eyes diverge when shifting focus from a near to a far object, and converge when focusing from a far to a near object. {\bf (b)} Our system can distinguish different vergence shifts. Each shift corresponds to a unique EOG pattern that can be classified by the proposed algorithms.}
    \label{fig:fig1}
\end{teaserfigure}


\begin{abstract}

There is growing industry interest in unobtrusive designs for electrooculography (EOG) sensing of eye gestures on glasses (e.g. JINS MEME and Apple eyewear). We present {\sysname}, an EOG-based glasses \orange{system} that enables depth-aware eye interaction \orange{by sensing vergence} with a \orange{glasses-compatible electrode layout} and smart glass prototype. \modify{It can distinguish between four depth-based eye gestures with 97\% accuracy on unseen users without any calibration in a user study across 20 users and 1,520 gesture instances.} 
To reduce false detections, we incorporate a motion artifact detection pipeline and a preamble-based activation scheme. The system uses dry sensors without any adhesives or gel and operates in real time with 3~mW power consumption by the analog sensing front-end.


\end{abstract}

\maketitle
\section{Introduction}

\begin{figure*}[h]
    \centering
    \includegraphics[width=1\linewidth]{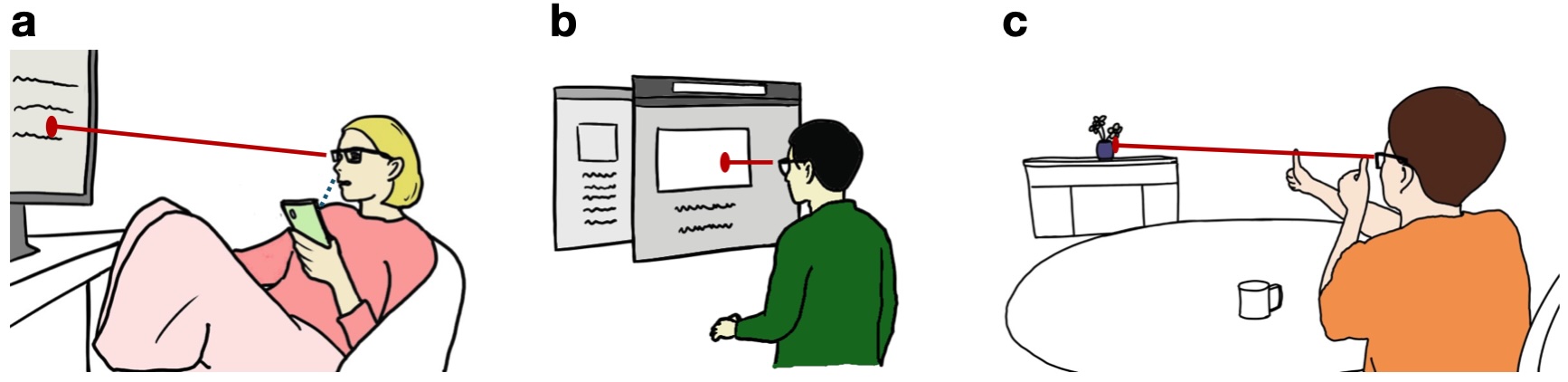}
    \caption{{\bf Applications of {\sysname}. (a)} Varifocal lenses adjust focal depth as the user shifts gaze between ambient reference points: a nearby phone and a distant TV. {\bf (b)} Smart glasses enable gaze-based selection between virtual objects at different depths. {\bf (c)} Remote screening of ocular disorders using approximate reference points: thumbs and a distant object.}
    \label{fig:applications}
\end{figure*}

Given their proximity to the eyes, glasses are uniquely positioned to leverage eye movements for interaction. A less-explored movement is vergence: the inward or outward motion of the eyes that occurs naturally as humans shift their gaze between objects at different depths (Fig.~\ref{fig:fig1})~\cite{Hughes1991Dic}. Unlike fast, reflexive saccades or lateral gaze shifts, vergence is slower and more deliberate, making it well-suited for intentional, hands-free interaction~\cite{ahn2020verge, kirst2016verge, zhang2024focusflow}. 


Glasses that can sense and respond to these depth-related eye movements can enable new forms of hands-free interaction across key depth cues (Fig.~\ref{fig:applications}):
\squishlist
\item \textit{Ambient depth cues.} Varifocal glasses can dynamically refocus between a nearby book and a distant TV, detect abrupt gaze depth shifts and \add{ flag attention lapses.} 
\item \textit{Virtual depth cues.} Augmented reality glasses can leverage vergence to select or focus on different virtual windows.
\item \textit{Approximate physical depth cues.} Remote screening for ocular disorders (e.g., convergence insufficiency, strabismus~\cite{mishra2020soft}) can be performed using one's thumbs and a distant object as approximate depth markers.
\squishend

Electrooculography (EOG) is a promising modality for vergence (gaze depth) sensing on glasses, with growing industry interest in integrating EOG electrodes into glasses. Notably, these sensors have already been integrated into the nose pads and bridge of commercial-grade glasses including JINS MEME~\cite{jinsmeme}, and more recently, Apple was granted a patent~\cite{patent} to embed EOG sensors into glasses. 

Unlike cameras, EOG is robust to varying ambient lighting conditions as glasses are worn across a range of environments. EOG is also low-power compared to commercial-grade cameras. In contrast to mmWave~\cite{ma2025mmet} and acoustic~\cite{li2024gazetrak} systems that place sensors in front of the lenses and can obstruct the user's vision, EOG electrodes have been integrated unobtrusively into the frames of the glasses. It is for these reasons that prior work~\cite{scharer2025electrasight, kosmyna2019attentivu2, frey2024gapses, Lee20203DPrinted, kim2024privacy, rostaminia2019w} has investigated EOG for capturing eye movements such as blinks and coarse gaze shifts (e.g., left, right, up, down) to support the understanding of interaction intentions.

\add{The challenge, however, is that current EOG systems for glasses are not designed to robustly detect vergence. During vergence, the eyes move in opposing directions to adjust viewing depth, which creates \textit{opposing electrical signals} that can interfere and become difficult to detect at the sensing electrodes~\cite{uema2017jins}.}

We present {\sysname}, an EOG-based glasses \orange{system} that expands the range of detectable eye movements to include vergence gestures. \orange{We identify two design objectives for robust vergence sensing: preserving the separability of signal contributions from the left and right eyes to limit interference, and maintaining wide sensor spacing to support a high signal-to-noise ratio (SNR). Accordingly,} we \orange{use} a two-channel design that records signals from the left and right eyes on separate channels using a shared reference electrode on the nose bridge and electrodes on the temples. \orange{We further implement this configuration in a hardware prototype compatible with the form factor of everyday glasses.}

Using this hardware setup, we show that our system can classify \add{four to six} vergence gestures that produce distinguishable EOG signals. These gestures involve changing gaze depth between three distances that span everyday viewing distances: reading distance (30~cm), social interaction distance (70~cm), and far-field visual attention (200~cm).


\add{However, in real-world use, deploying precise depth references is not practical, so our system evaluates two strategies for ad hoc depth references on glasses.} \textbf{For glasses with displays} such as augmented reality smart glasses, our system can render distance markers using on-screen visual overlays. Although these markers are displayed on a fixed physical screen, they simulate depth using binocular disparity, which causes users to exhibit vergence when gazing between different depths. \textbf{For glasses without displays}, users can use their outstretched thumbs as \textit{approximate} depth references: one at full arm's length ($\approx$70~cm) and the other at half arm's length ($\approx$30~cm), for example resting on the opposite elbow, along with any arbitrary far-field target. \add{This approach extends to basic audio-only frames like Amazon Echo Frames~\cite{amazon_frames} or instrumented varifocal glasses.}



\noindent Our contributions are as follows:
\squishlist
\item \add{We present a depth-aware glasses system that distinguishes four vergence gestures with 97.01\% accuracy when evaluated on unseen users without calibration in an IRB-approved study with $N = 20$ users.} The system is also compatible with traditional EOG sensing gestures including blinks and coarse gaze shifts and achieves 96.10\% accuracy across an 11-class set in a feasibility evaluation.
\item A smart glasses prototype to sense vergence with flexible electrode contacts to accommodate different users. The prototype has an analog front-end power consumption of 3~mW and can run in real-time on a lightweight microcontroller. The total weight of our system is 75~g.
\item To \orange{support} robust performance in real-world scenarios, we design a motion artifact detection pipeline to discard unwanted events caused by facial and body movements. \modify{{\sysname} achieved an accuracy of 98.77\% with a false rejection rate of 1.08\% and a false positive rate of 1.23\% on a dataset collected from $N = 6$ participants.}
\item We design a preamble-based activation scheme, where a brief eyebrow raise is performed to activate or deactivate vergence sensing. This reduces the false positive rate across several activities from \modify{0.50--2.43\% to nearly zero}, and ensures that depth-based interactions are only initiated when intended.
\item We open-source our hardware design and software code\footnote{ \url{https://github.com/zx-explorer/VergeIO}} for the community to build on.
\squishend

We envision that the sensing capability introduced by our approach can broaden the scope of eye interaction available on EOG-based glasses, enabling applications in consumer, health, and safety-monitoring domains.


\section{Background and related work}

\begin{table*}[t]
\centering
\begin{tabular}{|l|l|c|c|c|c|l|}
\hline
\textbf{Reference} & \makecell{{\bf Sensing}\\{\bf Modality}} & \makecell{{\bf Vergence}\\{\bf Sensing}} & \makecell{{\bf Robust to}\\{\bf Ambient Light}} & \makecell{{\bf Unobstructed}\\{\bf View}} & \makecell{{\bf Glasses}\\{\bf Form Factor}} & \makecell{{\bf Power}} \\
\hline
CIDER~\cite{mayberry2015cider} & Cameras & \cmark & \xmark & \xmark & \cmark & 40~µW \\ \hline
Tobii~\cite{tobii} & Cameras & \cmark & \xmark & \cmark & \cmark & DNS \\ \hline
FocusFlow~\cite{zhang2024focusflow} & Cameras & \cmark & \cmark & \cmark & \xmark & DNS \\ \hline
Li et al.~\cite{li2018battery} & Photodiodes & \cmark & \xmark & \cmark & \cmark & 395~µW \\ \hline
GazeTrak~\cite{li2024gazetrak} & Acoustic & \cmark & \cmark & \xmark & \cmark & 16.4~mW \\ \hline
mmET~\cite{ma2025mmet} & mmWave & \cmark & \cmark & \xmark & \cmark & 6~mW \\ \hline
JINS MEME~\cite{jinsmeme} & EOG & \xmark & \cmark & \cmark & \cmark & DNS \\ \hline
Mishra et al.~\cite{mishra2020soft} & EOG & \cmark & \cmark & \cmark & \xmark & DNS \\ \hline
ElectraSight~\cite{scharer2024electrasight} & EOG (hybrid) & \xmark & \cmark & \cmark & \cmark & 3.3~mW \\ \hline
Lee et al.~\cite{Lee20203DPrinted} & EOG & \xmark & \cmark & \cmark & \cmark & DNS \\ \hline
GAPses~\cite{frey2024gapses} & EOG/EEG & \xmark & \cmark & \cmark & \cmark & 6.2~mW \\ \hline
\rowcolor{gray!15}
{\bf {\sysname} (ours)} & {\bf EOG} & \cmark & \cmark & \cmark & \cmark & {\bf 3~mW} \\
\hline
\end{tabular}
\caption{Comparison of existing eye sensing systems on glasses. {\sysname} is explicitly designed for vergence sensing using an EOG electrode configuration for glasses. (DNS = did not specify)}
\label{tab:vergence_comparison}
\vspace{-2em}
\end{table*}


\subsection{Eye movement sensing on headworn devices}
\label{sec:related_eye_movement}
There has been increased interest in eye movement sensing on head-mounted devices. Camera-based approaches~\cite{mayberry2015cider,mayberry2014ishadow,pupil,tobii,cho2012gaze,ryan2008limbus}, while accurate, are power-hungry and are typically ineffective in the dark without infrared illumination. Photodiode-based systems~\cite{li2018battery}, though lower in power, can be saturated by ambient sunlight in outdoor settings. Emerging acoustic~\cite{li2024gazetrak} and mmWave systems~\cite{ma2025mmet} offer a promising low-power alternative, but can obstruct the user's line of sight.

Electrode-based approaches have made their way into VR headsets~\cite{Pai2022NapWell}, \modify{goggles~\cite{bulling2009wearable}, smart glasses~\cite{frey2024gapses, kosmyna2019attentivu2, Kanoh2015Jinsmeme, rostaminia2019w, Kosmyna2024WearablePair, Lee20203DPrinted, scharer2024electrasight, kim2024privacy}, and earable devices~\cite{king2025eareog, manabe2013conductive, liu2025eareog, lepold2024openearable}}. They are often able to simultaneously record electroencephalography (EEG), electrooculography (EOG), and electromyography (EMG) signals. 

Dry electrodes have been integrated into several commercial smart‐glasses platforms such as JINS MEME~\cite{jinsmeme} and AttentivU~\cite{kosmyna2019attentivu2} due to their small size, weight, and low power-consumption. They have achieved reliable blink detection and horizontal/vertical saccade tracking~\cite{uema2017jins}, facial‐expression recognition~\cite{rostaminia2019w}, midair gesture and context recognition~\cite{Yeo2021Jinsense}, fatigue estimation~\cite{Tag2019Jinsmeme}, and mediated attention monitoring with real‐time biofeedback~\cite{Kosmyna2018AttentivU}. We build on this rich body of prior work by enabling vergence and depth-awareness as a key interaction primitive for EOG-based glasses.

\subsection{EOG-based sensing systems}
\label{sec:related_eog_sensing}

EOG has been used to detect various eye movements, including vergence, saccades, blinks, gaze shifts, and fixations ~\cite{belkhiria2022eog,barea2002system}. However, it is important to note that EOG is generally not suitable for high-precision tracking. \orange{Despite a spatial resolution limited to approximately 1.5°~\cite{Young1975}, Stevenson et al.~\cite{stevenson2015estimating} demonstrated fixation depth estimation using an eight-channel wet-electrode EOG setup, reporting an average fixation distance error of $10.3 \pm 10.0\%$ ($5.7 \pm 4.7$ cm). This suggests that obtaining stable absolute depth estimates remains challenging.} 

Prior systems have leveraged EOG-detected eye movements as an input for assistive interfaces for users with physical disabilities~\cite{patmore1998towards}, robotic control~\cite{chen2004human}, and hands-free operation of wearable computers for caregivers~\cite{mizuno2003development}. Beyond eye movement, EOG has also been used to detect facial expressions (e.g. movements of the cheek, brow, and nose)~\cite{rostaminia2019w} and to recognize activities such as typing, reading, eating, and talking~\cite{ishimaru2014smarter}.

As EOG only requires small dry electrodes, there has been growing interest to incorporate them into glasses with academic prototypes including  GAPses~\cite{frey2024gapses}, E-Glasses~\cite{Lee20203DPrinted} and ElectraSight~\cite{scharer2024electrasight}, and notable commercial efforts including JINS MEME~\cite{jinsmeme}, and Apple's recently approved patent~\cite{patent} to incorporate EOG electrodes into glasses. However, currently proposed electrode configurations are either not able or not focused on sensing vergence, and are limited to blinks, winks, saccades and directional gaze shifts; or require a large number of electrodes that can be uncomfortable. In contrast, {\sysname} expands the set of eye movements that can be detected on glasses to include four to six different vergence gestures using \orange{a glasses-compatible EOG configuration designed to} \modify{preserve the opposing electrical polarity signals produced by vergence.}

\subsection{Vergence as an interaction gesture}
Vergence has emerged as a reliable eye-based input modality in extended reality environments~\cite{ahn2020verge, zhang2024focusflow} due to its ability to mitigate the ``Midas touch'' problem, which refers to the difficulty of separating deliberate gaze commands from involuntary natural exploration as the eyes move around a scene~\cite{Jacob1990Midas}. Prior work has used eye-tracking cameras to measure vergence in VR systems for selecting and activating virtual objects~\cite{ahn2020verge,zhang2024focusflow}, controlling linear actuators~\cite{ruan2018human}, and studying visual attention in children with reading difficulties~\cite{jimenez2020eye}. EOG-based systems include configurations that are not compatible with a glasses form factor~\cite{Tringali2021eyeaccommodation} and customized flexible-electrode designs that require specialized fabrication~\cite{mishra2020soft}, making them more challenging to scale. In contrast, our approach focuses on detecting vergence using an EOG-based electrode configuration for glasses using conventional Ag/AgCl dry electrodes.


\section{System design}


\subsection{Vergence distance selection}
\label{sec:distance_selection} \orange{Given the spatial-resolution limitations of EOG discussed in Sec.~\ref{sec:related_eog_sensing}, }we define the depth-aware task for EOG as detecting transitions between discrete distances rather than performing continuous depth regression. Therefore, a key system design parameter is the selection of vergence distances which is the physical distance between the eye and a fixation distance. These distances must satisfy two requirements: First, they should span the range of realistic, everyday viewing distances, and second, changes in vergence between these distances should result in discernible EOG signals that can be classified.

\begin{figure}[H]
    \centering
    \includegraphics[width=1.0\linewidth]{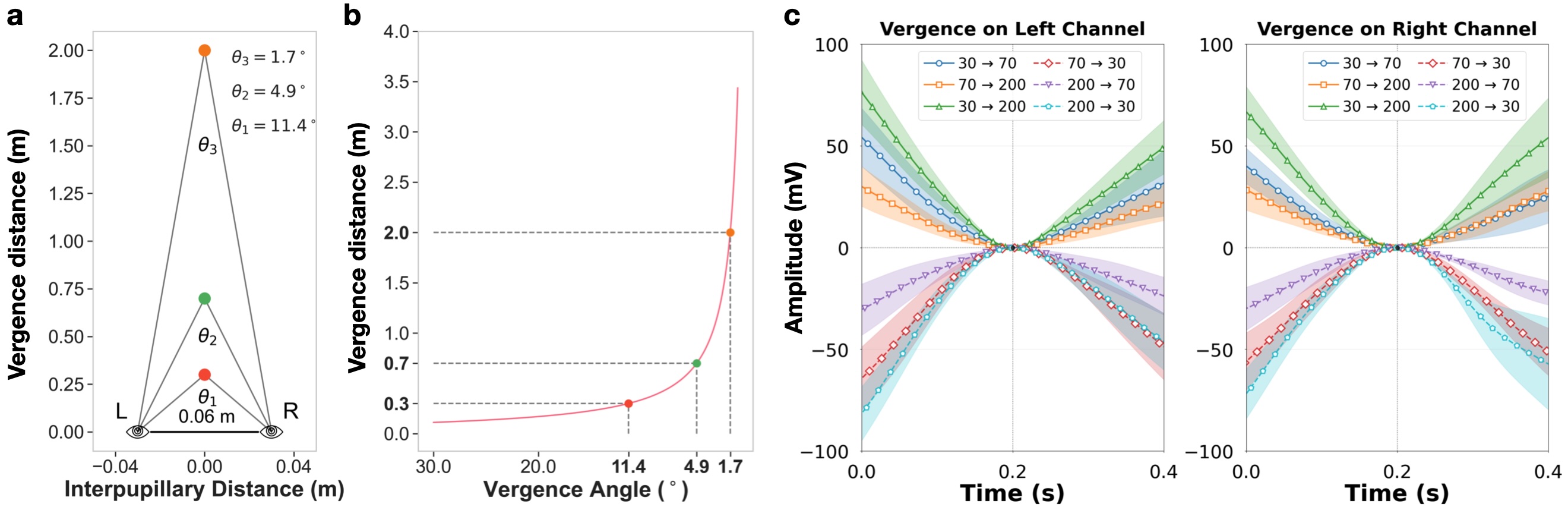}
    \caption{\textbf{Geometric characterization and signal validation for vergence distance selection.} \textbf{(a)} {\sysname} measures \textit{changes} in vergence angle ($\theta$) using EOG signals, which vary proportionally with these changes. \orange{Interpupillary distance of 60~mm is used for illustration.} \textbf{(b)} Our system can distinguish changes in vergence angles corresponding to three interaction distances: 30, 70, and 200~cm. \textbf{(c)} Mean EOG signal for each of the six vergence gestures for the left and right eye, \orange{with each trace’s extremum aligned to (0.2~s, 0~mV) for comparison.} Shaded region represents one standard deviation from the mean.}
    \label{fig:depth-analysis}
\end{figure}

Since EOG measures eye movements and directly reflects \textit{changes in vergence angle}, we choose distances such that the resulting changes in vergence angle are uniformly spaced. Fig.~\ref{fig:depth-analysis}a shows how vergence angle ($\theta_1$, $\theta_2$, $\theta_3$) is defined as the angle between the lines of sight of both eyes as they converge on a fixation point.

\orange{Binocular vergence angle for a target at distance $d$ is given by $\theta(d) = 2\arctan\left(\frac{\mathrm{IPD}}{2d}\right)$. Within our interaction range ($d \geq 30$~cm, according to ISO 9241-303~\cite{iso9241}) and an upper-bound adult IPD of $75$~mm~\cite{ipd_quest,dodgson2004variation}, $\frac{\mathrm{IPD}}{2d} \leq 0.125 \ll 1$. Therefore, $\arctan(x) \approx x$, yielding $\theta(d) \approx \frac{\mathrm{IPD}}{d}$. As an individual's IPD is fixed, the vergence shift between depths simplifies to $\Delta\theta_{ij} \approx \mathrm{IPD}\left\vert{}\frac{1}{d_i}-\frac{1}{d_j}\right\vert{}$. We therefore select viewing distances of $30$, $70$, and $200$~cm, which produce pairwise vergence angle changes ($\Delta\theta$) in an approximate 1:2:3 ratio.} \add{As EOG responses can be regarded linear for gaze angles within $\pm$ 30°~\cite{Estrany2022Human}, our selection increase the separability among all possible vergence gestures in practice.}

\modify{As shown in Fig.~\ref{fig:depth-analysis}b, because the relationship between vergence angle and distance is non-linear, these vergence angles correspond to three vergence distances at key interaction distances: near (30~cm), mid (70~cm), and far (200~cm), which align with focal zones outlined by the American Optometric Association (AOA)~\cite{mancil2011optometric}.}

We show in Fig.~\ref{fig:depth-analysis}c the mean EOG curves for each of the six gestures when performed using our proposed electrode placement by one participant. \add{The plot shows that the direction of depth shift is separable due to EOG signals with opposite polarities generated by convergence and divergence. For gestures moving in the same depth direction, the largest angular shift (30 $\leftrightarrow$ 200~cm) and the smallest angular shift (70 $\leftrightarrow$ 200~cm) are visually distinct. However, we do note 30 $\leftrightarrow$ 70~cm transitions exhibit overlap with other gestures, making them challenging to separate visually. This motivates our decision to focus on a 4-gesture set as the primary use case for our system.}




\subsection{Placement of EOG electrodes}
\label{sec:electrodes_placement}

Electrode placement is a critical consideration in our design as it directly impacts the types of eye movements that can be detected. \add{Existing EOG sensing systems on glasses generally rely on two design paradigms. The first employs compact electrodes around the nose pads and bridge, as represented by commercial systems like JINS MEME~\cite{jinsmeme} and academic prototypes such as GAPses~\cite{frey2024gapses}. The second relies on wider sensor placements near two temples, as represented by E-Glasses~\cite{Lee20203DPrinted}. However, neither paradigm is specifically designed to effectively capture vergence.}

\begin{figure*}
    \centering
    \includegraphics[width=\linewidth]{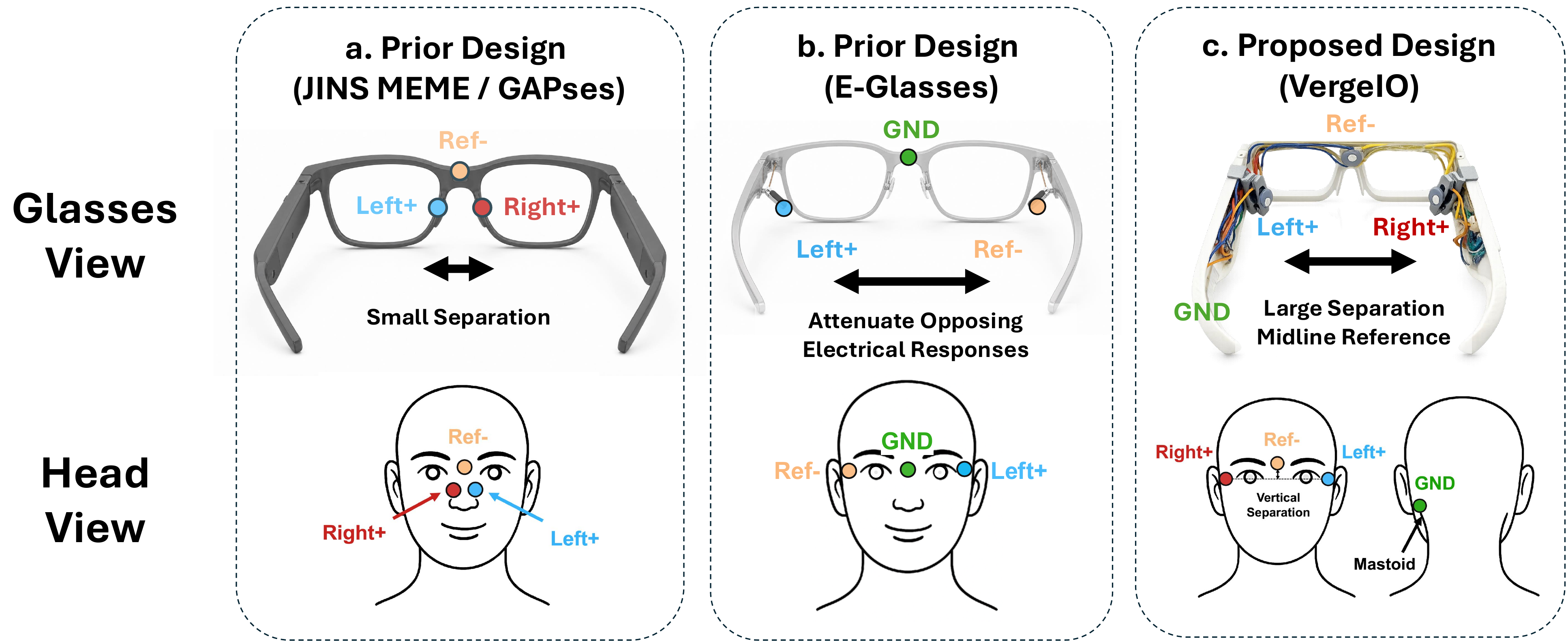   }
    \caption{\add{{\bf \orange{Representative glasses-based EOG electrode configurations.} (a)} \orange{Compact nose-pad and bridge configuration represented by JINS MEME~\cite{jinsmeme} and $\text{GAP}{\text{ses}}$~\cite{frey2024gapses}.}
    {\bf (b)} \orange{Wider temple configuration represented by E-Glasses~\cite{Lee20203DPrinted}, which records a single differential channel across the two sides.} {\bf (c)} {\sysname} combines wide sensor separation with a shared nose-bridge reference, capturing vergence with high SNR while preserving the ability to sense non-vergence eye movements.}}
    

    \label{fig:electrode_placement}
\end{figure*}

JINS MEME places both positive electrodes on the nose pads and the negative electrode on the nose bridge with minimal spatial separation (Fig.~\ref{fig:electrode_placement}a). This compact configuration primarily captures conjugate changes \add{(e.g. blinks, saccades)} rather than the asymmetric signals required to detect vergence. With limited spatial separation, the opposing electrical dipoles generated during vergence tend to interfere, cancel out, or simply become indistinguishable. This limitation was explicitly confirmed by prior work authored by JINS MEME researchers~\cite{uema2017jins}.

\add{E-Glasses~\cite{Lee20203DPrinted} (Fig.~\ref{fig:electrode_placement}b) increases spatial separation by placing electrodes near two temples \orange{and records a single differential channel across the two sides}, which is effective for sensing conjugate movements (e.g., saccades) where eyes rotate in the same direction. However, the two eyes rotate in opposite directions during vergence, resulting in the electrical potentials at both temples rising or falling together, which can result in partial cancellation. This limitation was confirmed by Mishra et al.~\cite{mishra2020soft} in a vergence classification task, where a conventional differential setup achieved only 34\% accuracy.}


To address these limitations, our system is designed to support vergence sensing while being compatible with a glasses form factor. Our design (Fig.~\ref{fig:electrode_placement}c) places positive electrodes at each temple (left and right channels), a shared negative electrode on the nose bridge, and a ground electrode at the mastoid. \modify{This configuration offers a twofold advantage: the lateral sensing electrodes provide sufficient spatial separation to capture vergence responses with high SNR, while the shared nose-bridge reference preserves the opposing-polarity signals generated by convergence and divergence.} Additionally, by introducing a vertical separation between the temple and nose-bridge electrodes, our design preserves the ability to sense non-vergence eye movements (e.g., saccades and blinks). Together, {\sysname} expands vergence sensing capability while maintaining compatibility with conventional horizontal and vertical EOG-based eye-movement sensing, which was validated in Sec.~\ref{sec:11way}.




\orange{To compare the signal quality of our approach with the configurations discussed above}, we conduct a pilot study to measure the signal-to-noise ratio (SNR) of vergence gestures captured \modify{by {\sysname} against the two paradigms (JINS MEME~\cite{jinsmeme} and $\text{GAP}{\text{ses}}$~\cite{frey2024gapses}; E-Glasses~\cite{Lee20203DPrinted})}.  In this study, one participant simultaneously wore electrodes \modify{required for these layouts}: two electrodes on the temples, two on the nose pads, a shared electrode on the nose-bridge as the reference, and a ground electrode at the mastoid.


\begin{figure}
    \centering
\includegraphics[width=0.6\linewidth]{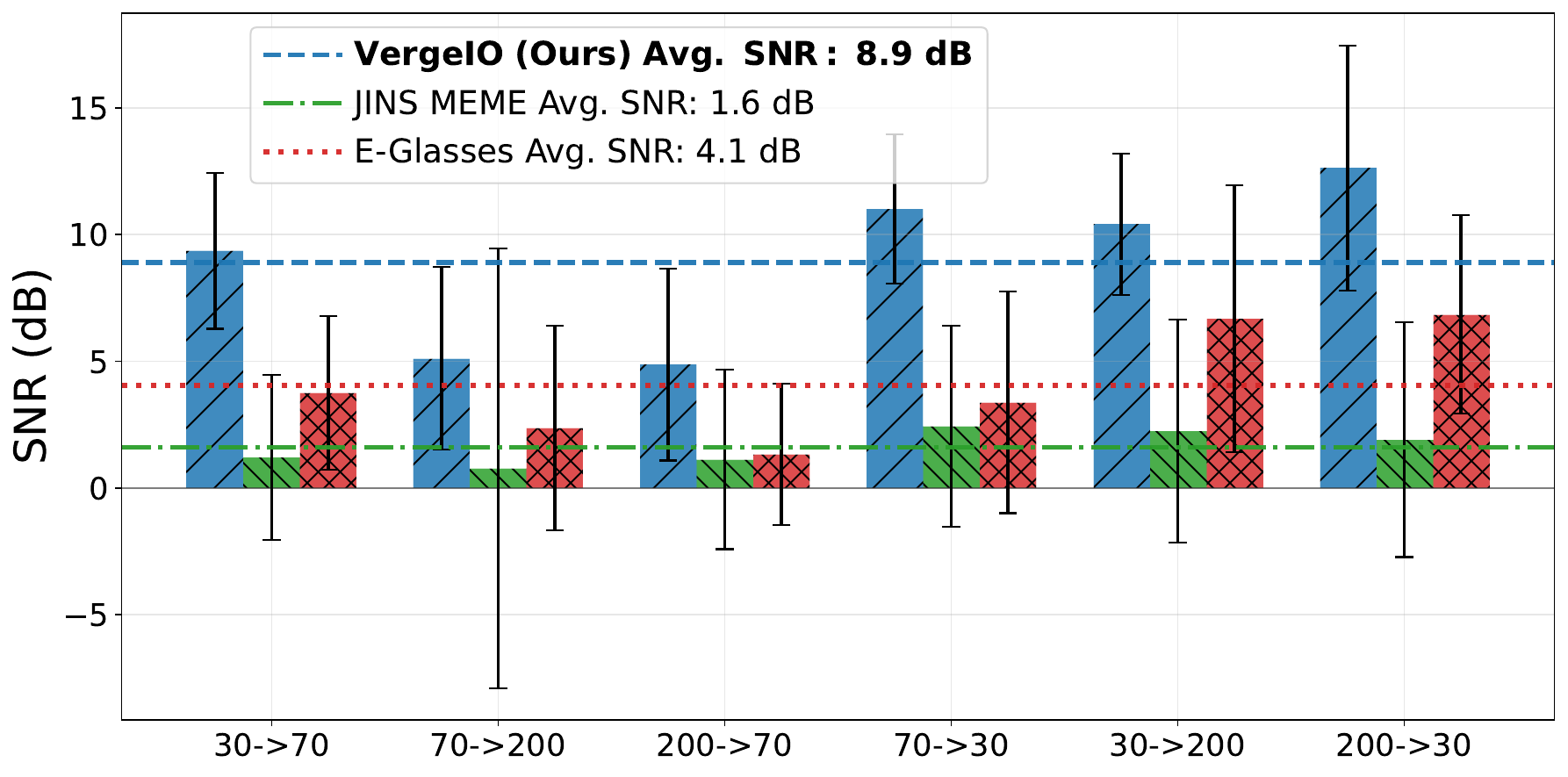}
\caption{\add{\textbf{SNR comparison across six vergence gestures.} \orange{In this pilot study, }{\sysname} outperforms the electrode layouts from prior designs in JINS MEME glasses~\cite{jinsmeme} and E-Glasses~\cite{Lee20203DPrinted}. We show the left channel for {\sysname} and JINS MEME. E-Glasses outputs a single channel only.}}
\label{fig:jinsmeme_comparison}
\end{figure}

We recorded four-channel EOG data over ten rounds, where each round consisted of six vergence gestures between the three distances we selected in Sec.~\ref{sec:distance_selection}. Synchronized data were sampled at 500 Hz and processed in sequence with a third-order zero-phase low-pass filter at 10 Hz, a notch filter, and a third-order Savitzky–Golay filter~\cite{savitzky1964smoothing} (window size is 0.5 second). We manually segmented vergence events from the filtered signals of our two channels. Since distinguishable event boundaries were not visible in the JINS MEME channels, we applied the onset and offset times identified from the {\sysname} channels to all configurations. For each event, we extracted the two 0.2-second signals immediately before and after the event as noise segments. SNR was then computed as the ratio of the root-mean-square amplitude of the event segment to that of the concatenated noise segments.

\modify{As shown in Fig.~\ref{fig:jinsmeme_comparison}, the average SNR across all six vergence gestures was 8.9~dB with our system, compared to 1.6~dB with the JINS MEME layout and 4.1~dB with the left-right differential layout represented by E-Glasses.} \orange{These results indicate stronger EOG responses during vergence with our system.} 





We note that while prior work~\cite{narayan2024eog} on EOG sensing incorporates vertical electrodes above and below each eye, vergence is primarily a horizontal movement, and vertical channels only provide modest performance gains. Additionally, we found in our evaluation that (a) participants reported that the additional vertical electrodes caused discomfort (Sec.~\ref{sec:vergence_classification_evaluation}) and (b) our electrode configuration is still able to classify between different vertical eye movements (Sec.~\ref{sec:11way}).

\begin{figure*}
    \centering
    \includegraphics[width=1\linewidth]{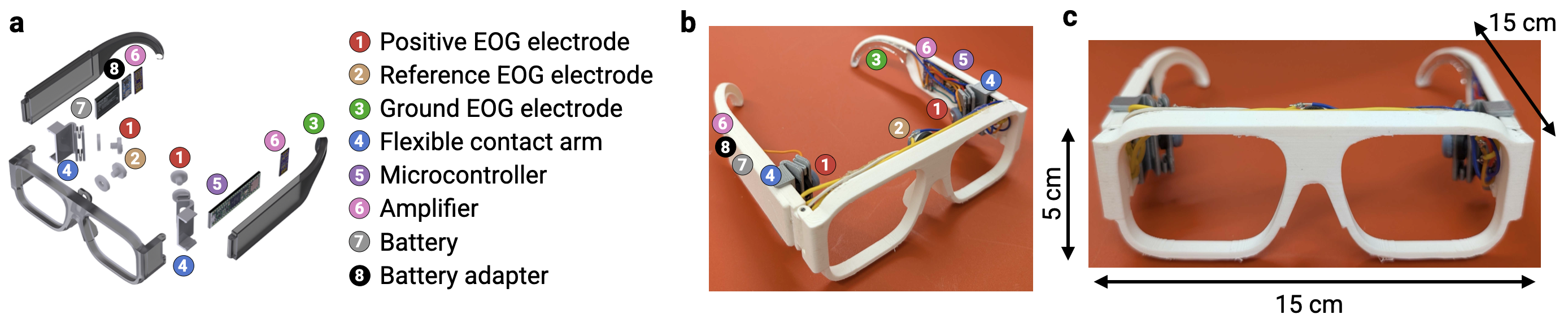}
    \caption{{\bf Hardware design. (a)} Exploded CAD design illustrating the different components of the hardware design. {\bf (b)} Fabricated design annotated with location of different components. {\bf (c)} Front-view of glasses with dimensions.}
    \label{fig:hardware_design}
\end{figure*}

\subsection{Hardware design}
\label{sec:hardware_design}

We create a glasses hardware prototype to ensure reliable skin contact of our chosen electrode placement (Fig.~\ref{fig:hardware_design}):
\squishlist
\item {\bf Electrodes.} We use dry electrodes for practicality, similar to other glasses with biosignal electrodes~\cite{jinsmeme,kosmyna2019attentivu2}. We selected SoftPulse~\cite{Chen2014Softpulse} which has been used for EOG acquisition~\cite{Chen2014Softpulse, frey2024gapses, Heijden2024MultiChannel}, and is biocompatible with the skin (ISO 10993-5/10).

\squishlist
\item \textit{Positive electrodes.} These are coupled to flexible contact arms at the user's temple for good skin contact. These two arms can slide along the glasses’ temples to accommodate different users.

\item \textit{Reference electrode.} This is mounted at the nose bridge to ensure direct contact with the user’s skin. 

\item \textit{Ground electrode.} Curved coating at the end of the arm to maintain good contact with the user’s mastoid.
\squishend
\item {\bf Amplifiers.} We chose the BioAmp-EXG-Pill~\cite{amplifier} for its compact analog front end (2.5 $\times$ 1 $\times$ 0.1~cm) with a 1000× signal gain, and its ability to interface with any 3.3 and 5~V MCU via ADC.
\item {\bf Microcontroller.} We selected the Teensy 4.1~\cite{teensy}, which integrates 12-bit ADC and high-performance Cortex-M7 core, enabling precise, high-resolution bio-signal digitization and real-time processing.
\item {\bf Battery.} We used the same 570~mAh rechargeable lithium battery as in Google Glass~\cite{gglass}. As discussed in Section ~\ref{sec:power_consumption}, this battery enables sensing for approximately 197~hours (8.2 days). We also integrated the Adafruit Micro Lipo Charger 4410~\cite{adafruit} into the glasses to enable on-board charging of the sensing system.
\squishend

\noindent {\bf Cost.} The off-the-shelf cost is \$140, with the optimized cost \$73. The cost of the ELDRY electrode coating ($4\,\mathrm{cm}^2$, 0.5~mm thickness) is \$20; Li-Po rechargeable battery is \$15; Li-Po micro charger is \$6; microcontroller development board is \$30; and two BioAmp-EXG-Pill amplifier modules is \$69, which can be reduced to \$2 when using the discrete components.

\subsection{Motion artifact and noise removal}
\label{sec:noise}




Given that EOG signals are highly susceptible to facial and body movements~\cite{rostaminia2019w}, the identification and removal of such artifacts are imperative to prevent false positives. Since the resulting EMG (muscular) responses overlap with the EOG signals in frequency, they cannot be easily isolated in the frequency domain.

For this reason, we leverage a temporal, data-driven segmentation approach to identify and remove segments affected by user movement. Our noise removal pipeline is designed to operate on streaming EOG data, and operates on a moving window of 2~s with a step size of 0.1~s. We chose 2~s as it is ample time to perform a vergence movement, which typically takes less than 1~s~\cite{tyler2012analysis}.

\begin{figure}[h]
    \centering
    \includegraphics[width=\linewidth]{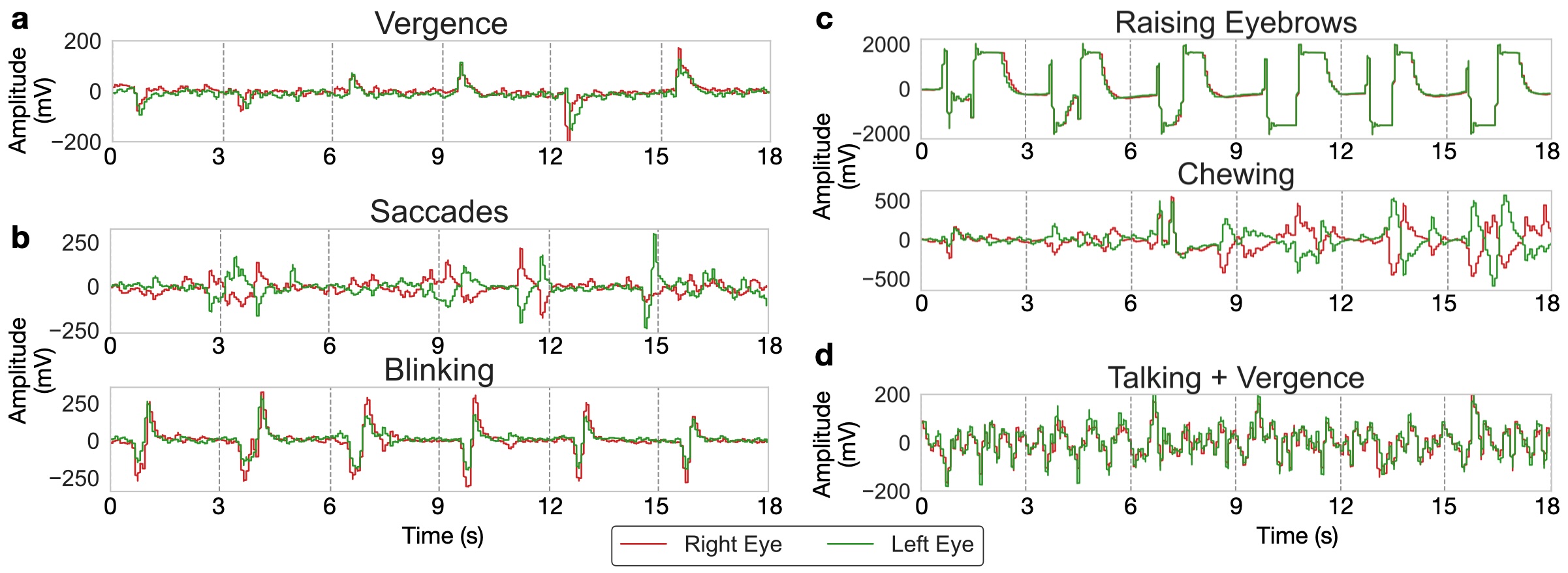}
    \caption{{\bf EOG signals recorded from the left and right eye during vergence, motion and noise events. (a)} Six vergence gestures. {\bf (b)} Vergence is distinct from eye movements like saccades and blinking. {\bf (c)} Facial movements produce higher amplitude signals than vergence alone. {\bf (d)} When combined with facial movements, vergence is dominated by facial activity, and is treated as noise.}
    \label{fig:segments}
\end{figure}

We collected a dataset (Table~\ref{tab:recordings}) comprising different motion and noise signals from six participants (Fig.~\ref{fig:segments}). \modify{Participants were instructed to perform {\bf (1)} vergence movements; {\bf (2)} body and head motions (e.g., turning, head tilt, standing/sitting); {\bf (3)} facial and eye movements typically considered noise (e.g., brow raising, saccades, blinking). Participants were cued with an audible beep every 3 seconds to perform every movement.}





\begin{table}[t]
\centering
\small
\begin{tabular}{|c|c|c|c|c|}
\hline
\textbf{Category} & \textbf{Description} & \textbf{\#Recordings} & \textbf{\#Events} & \textbf{Duration (min)} \\
\hline
Positive & Vergence & 60 & 360 & 25.20 \\
\hline
Noise & Blinking & 30 & 180 & 13.31 \\
Noise & Brow raising & 30 & 180 & 12.80 \\
Noise & Saccades & 30 & 180 & 12.99 \\
\hline
Motion & Head tilt & 30 & 180 & 12.83 \\
Motion & Head turn & 30 & 180 & 12.83 \\
Motion & Stand/Sit & 30 & 180 & 12.80 \\
\hline
\textbf{Total} & \textbf{All} & \textbf{240} & \textbf{1440} & \textbf{102.76} \\
\hline
\end{tabular}
\caption{\modify{Dataset collected from $N = 6$ participants for evaluating the artifact and noise removal pipeline.}}
\label{tab:recordings}
\vspace{-2em}
\end{table}


\begin{figure*}
    \centering
    \includegraphics[width=0.9\linewidth]{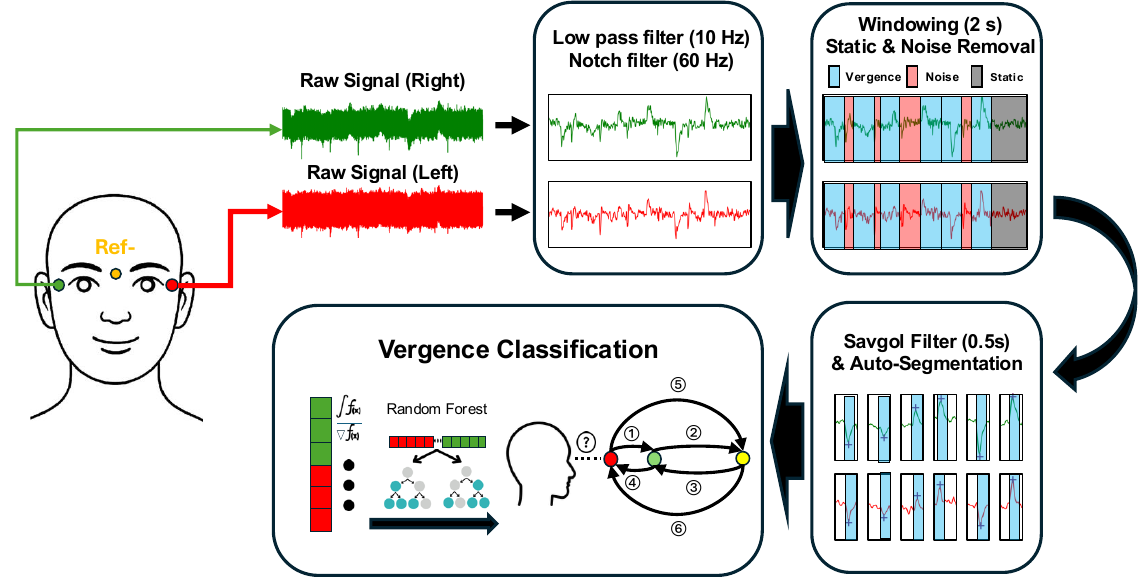}
    \caption{{\sysname} processing pipeline.
    }
    \label{fig:processing_pipeline}
\end{figure*}

\noindent {\bf Signal Conditioning.} All signals were preprocessed using a third-order low-pass Butterworth filter with a 10~Hz cutoff, which was chosen to attenuate high-frequency noise while preserving the dominant spectral content of the EOG signal~\cite{mishra2020soft}. A notch filter working at 60~Hz was also adopted to suppress powerline interference.

\noindent {\bf \add{Segmentation and} Silence Removal}. \modify{To support real-time processing of streaming EOG data, the filtered signal is segmented using a 2-s moving window with a 0.1-s stride.} We then apply a dynamic amplitude threshold to minimize computational overhead and prevent false positives from static noise, ensuring that only windows exceeding this threshold are processed. Specifically, the threshold is computed by applying a median absolute deviation (MAD) filter over the first 2 seconds of each channel, which are assumed to be static, and multiplying this by a heuristic value of 4.5.

\noindent {\bf Artifact Detection and Removal.} For each 2-second window we computed statistical features for both left and right EOG channels, including signal mean, extrema, band power, wavelet energy, variance, RMS, peak-to-RMS ratio, trapezoidal integral, maximum derivative, and minimum derivative. We trained a logistic regression classifier to distinguish between vergence and non-vergence (motion/noise) segments and remove the latter. We evaluated the pipeline in Sec.~\ref{sec:motion_eval}.


\begin{algorithm}
\caption{Vergence Gesture Classifier}
\begin{algorithmic}[1]
\setlength{\itemsep}{3pt}
\Procedure{ClassifyVergence}{$eog$}
  \State $L \gets []$
  \ForAll{$seg \in \text{Segment}(eog,2\text{s})$}
    \If{not \textsc{IsArtifact}($seg$)}
      \State $seg \gets \text{SavitzkyGolayFilter}(seg,250)$
      \ForAll{$p \in \text{DetectPeaks}(seg,30\text{mV})$}
        \State $g \gets seg[p\!-\!0.5,p\!+\!0.5]$
        \State $f \gets \text{ExtractFeatures}(g)$
        \State $l \gets \text{RandomForest.Predict}(\text{Z-Score Normalize}(f))$
        \State append $l$ to $L$
      \EndFor
    \EndIf
  \EndFor
  \State \Return $L$
\EndProcedure

\Function{ExtractFeatures}{$g$}
  \State Split $g$ into left and right halves; for each half compute
  \State\quad$[\max-\min,\;\int g,\;(g_e-g_s)/0.5,\;\overline{\Delta g},\;\mathrm{Var}(\Delta g)]$
  \State \Return concatenated 10‐dim vector
\EndFunction
\end{algorithmic}
\end{algorithm}

\subsection{Vergence classification pipeline}
\label{sec:classification_pipeline}


\noindent {\bf Automatic Segmentation.} After noise‐artifact removal, each 2~s signal window was smoothed using a Savitzky–Golay filter~\cite{savitzky1964smoothing} with a window length of 250 samples (0.5 s), which can clean the signal while preserving the features. Compared with moving averages or wavelet denoising, the Savitzky–Golay filter preserves edges more faithfully and is computationally efficient for real-time use~\cite{ACHARYA2016677, abdullah2023review}. To precisely determine the onset and offset of vergence gestures in each 2~s window, we employed the peak-detection algorithm in the detecta library~\cite{marcos_duarte_2021_4598962} with 30 mV as the minimum amplitude threshold, which we determined empirically from the pilot study in Sec.~\ref{sec:electrodes_placement}. Since vergence movements typically range less than 1s~\cite{Chen2014Softpulse}, we then extracted a window centered on the detected peak, extending 0.5 seconds before and after the peak, yielding a 1-second segment.

\noindent {\bf Feature Extraction.} Each 1~s segment can be regarded as two halves split by the peak we detected. From each half, we extracted five features: amplitude range, definite integral, end-to-end slope, mean of the first derivative, and variance of the first derivative. This results in ten features per EOG channel. Each gesture is represented by a 20-dimensional feature vector, comprising ten features extracted from each of the two channels. \add{Additional details on feature selection are provided in Appendix~\ref{sec:appendix_feature}.}


\noindent {\bf Vergence Classification.} Prior to model training, all features were standardized via z-score normalization to compensate for inter-user variability. The normalization also mitigates data drift due to shifts in electrode placement before and after eyeglass remounting. Using the PyCaret library~\cite{PyCaret}, we automated model selection and determined that the random forest was the most effective algorithm for classifying vergence gestures.
\section{Study design}
\label{sec:user_study_design}

\begin{figure}
    \centering
    \includegraphics[width=\linewidth]{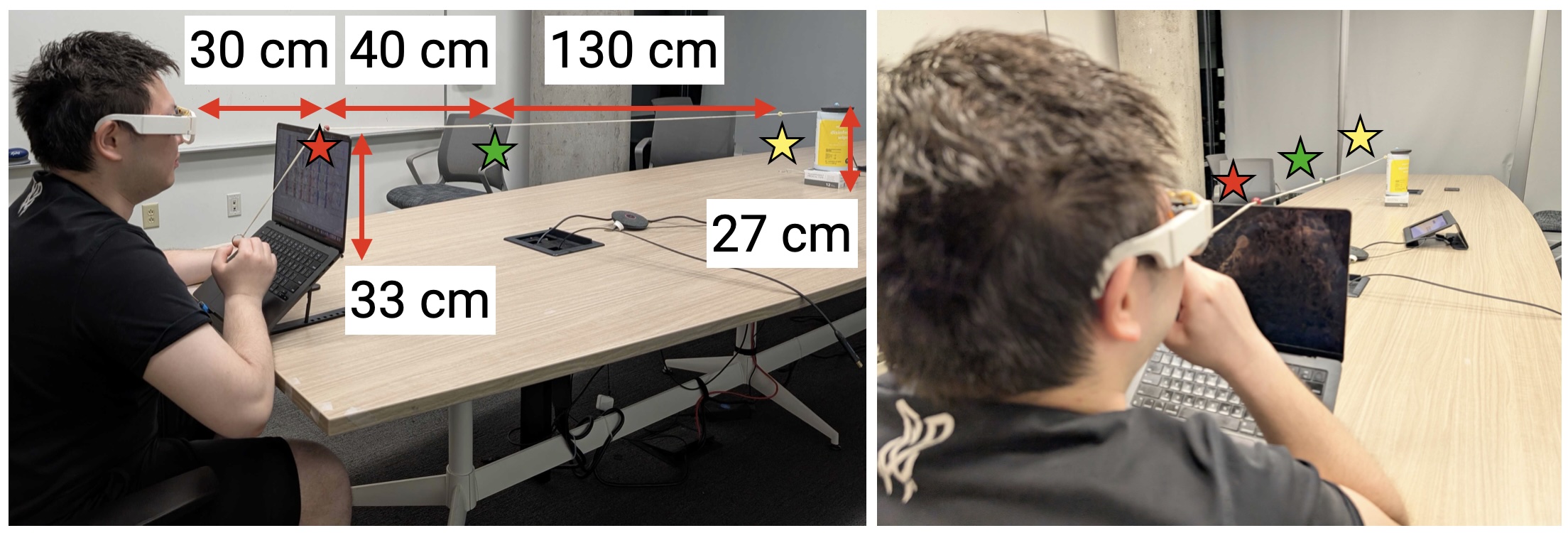}
    \caption{{\bf Experimental setup for user study.} The stars denote the location of colored beads tied to the Brock String and used as distance markers for the vergence gestures.}
    \label{fig:exp_setup}
\end{figure}

\begin{figure}
    \centering
    \includegraphics[width=0.96\linewidth]{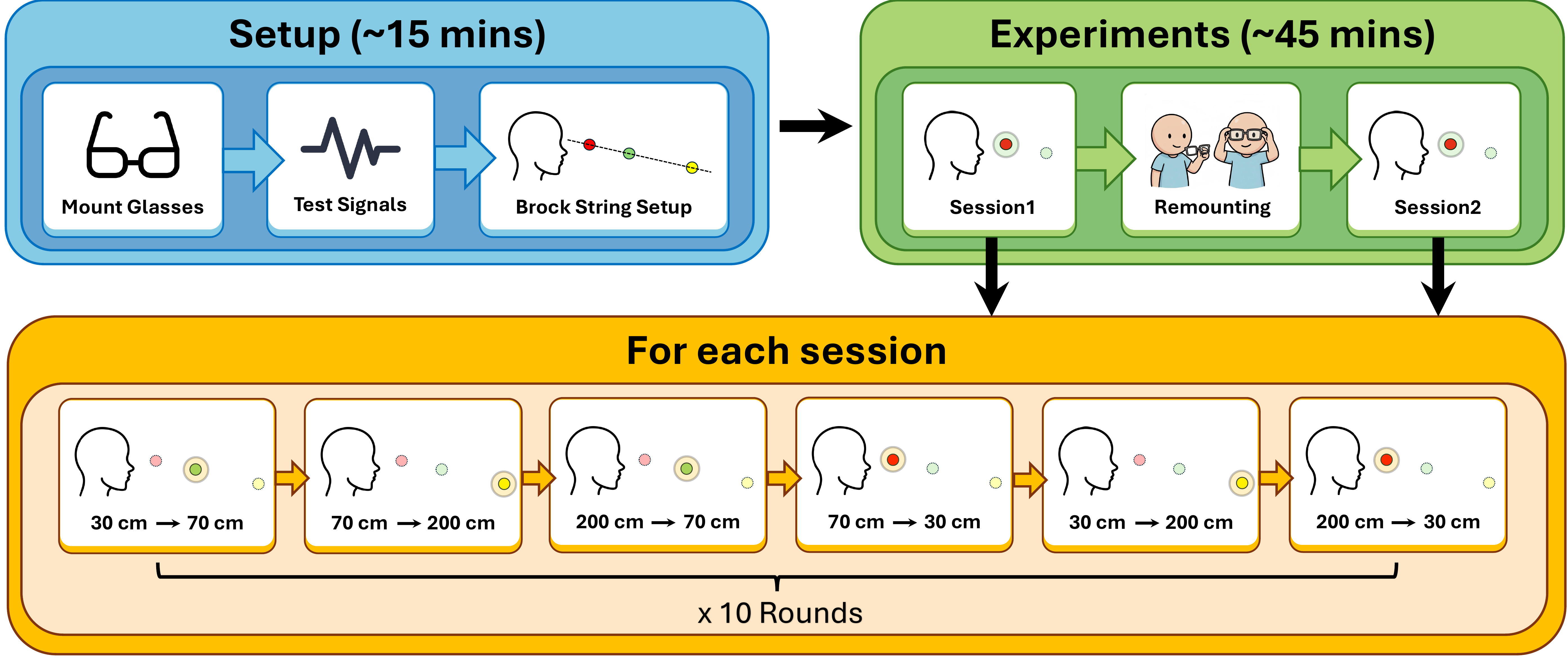}
    \caption{{\bf Experiment timeline.} Participants first underwent a setup phase to ensure the glasses were properly mounted and received training on performing vergence movements along the Brock String. During the experimental phase, they completed 10 rounds, each consisting of the six vergence gestures. After completing the rounds, participants removed and remounted the glasses before repeating the procedure.}
    \label{fig:timeline}
\end{figure}

\subsection{Experimental setup}
\label{sec:experimental_setup}

This study was approved by our Institutional Review Board (STUDY2025\_00000189). We recruited 20 participants (aged 25.75 $\pm$ 11.1), including 4 with normal vision and 16 with common refractive errors (12 myopia, 2 hyperopia, and 2 presbyopia). No participants reported other ocular-related disorders or convergence insufficiency.




All the experiments were conducted in a meeting room (Fig.~\ref{fig:exp_setup}). To trigger the six vergence gestures, we used a Brock String~\cite{hassan2017efficacy}, a clinical tool commonly used to assess binocular vision and diagnose convergence insufficiency. Our string includes three colored beads positioned at the designated vergence distances: red at 30~cm, green at 70~cm, and yellow at 200~cm.

In our setup (Fig.~\ref{fig:exp_setup}), participants sat in front of a laptop. The distal end of the Brock String was secured to a fixed anchor on the table at 27 cm above its surface, while the proximal end passes through a metal ring that the user holds taut at the bridge of their nose, positioned approximately 33~cm high to align with the participant’s line of sight. The string is tensioned such that it rests against the top edge of the laptop screen, helping to stabilize it throughout the task. Elevating the string in this way prevented any visual overlap of the beads, ensuring that each target remained clearly distinguishable.




\subsection{Data collection protocol}

We began by fitting each participant with our glasses prototype. The flexible contact arm was adjusted to ensure secure, comfortable electrode contact with the skin over the temples. We positioned the Brock String so that participants could maintain a forward gaze aligned with all three beads.

Following the Brock String exercise procedure~\cite{scheiman2008clinical}, participants were instructed to perform vergence gestures such that the target bead remained in sharp focus while the non-target beads appeared blurred. Our experimental protocol (Fig.~\ref{fig:timeline}) consisted of two sessions, each comprising ten rounds. In each round, participants performed all six vergence gestures once, cued by an audible beep, with a three-second window to complete each movement. After ten rounds, participants removed the glasses for a ten-minute break before remounting the device. For each participant, we disinfected all recording apparatus in accordance with institutional hygiene protocols.






\section{Evaluation}

\add{This section presents the evaluation results of {\sysname} under the controlled Brock String protocol.}

\subsection{Vergence classification}
\label{sec:vergence_classification_evaluation}

\begin{figure*}
    \centering
    \includegraphics[width=.98\linewidth]{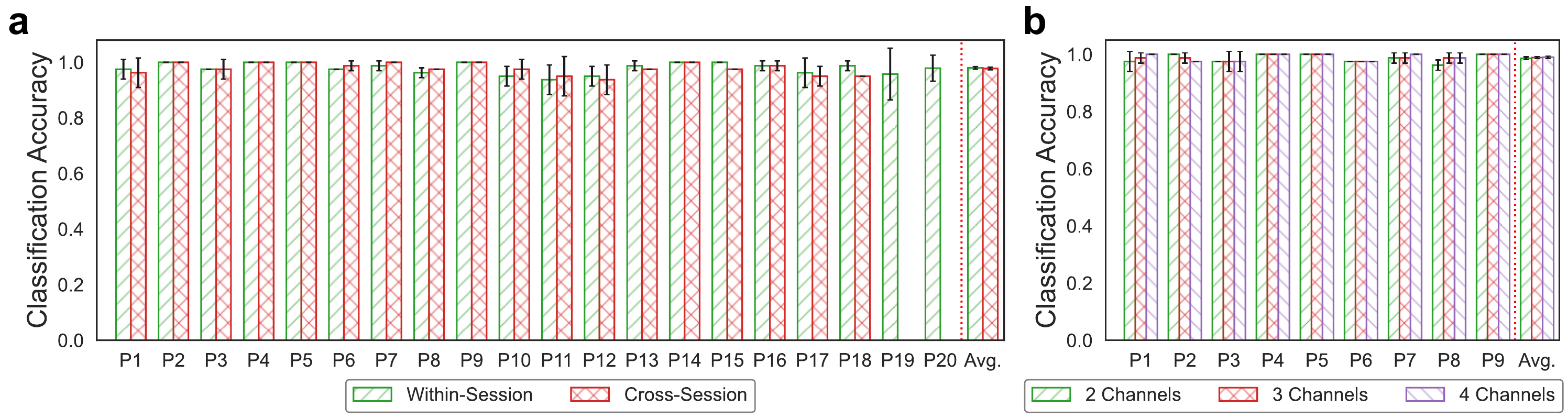}
    \caption{{\bf (a)} Classification accuracy ($n = 20$) for within-session (training and testing on data collected without remounting the glasses) versus cross-session (training on one session and testing on the unseen session after re-mounting glasses). P19 and P20 did not attend session 2. {\bf (b)} Within-session 5-fold cross-validation accuracy per participant ($n = 9$) for 2-, 3- and 4-channel configurations.}
    \label{fig:per-user-comparison}
\end{figure*}

\begin{figure*}
    \centering
    \includegraphics[width=0.96\linewidth]{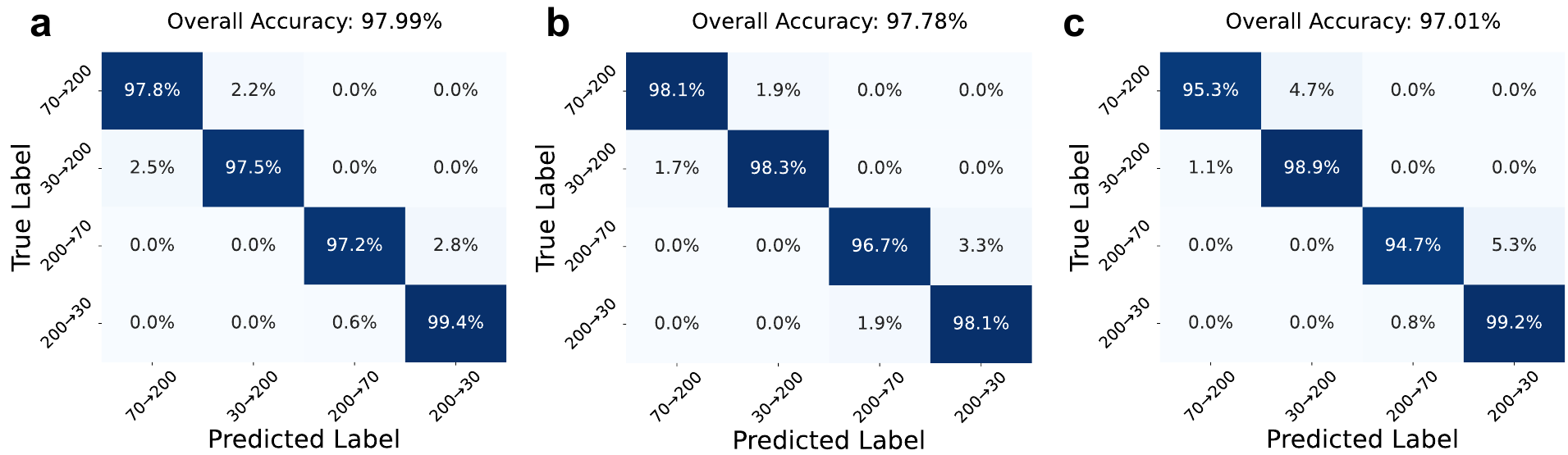}

    \caption{{\bf (a)} Confusion matrix of within-session 4 gestures classification across all the participants and sessions. {\bf (b)} Confusion matrix of cross-session 4 gestures classification. {\bf (c)} Confusion matrix of cross-user 4 gestures classification.}
    \label{fig:confusion_matrix}
\end{figure*}

\noindent {\bf Within-Session Performance.} We performed a within-session evaluation for each of the 20 participants in our study (Fig. \ref{fig:per-user-comparison}a, Table~\ref{tab:statistical_res}). \modify{Based on the observation in Sec.~\ref{sec:distance_selection}, we first evaluate our system on four gestures which have a high degree of separability: (30 $\rightarrow$ 200~cm, 70 $\rightarrow$ 200~cm, 200 $\rightarrow$ 30~cm, and 200 $\rightarrow$ 70~cm). {\sysname} achieved an average accuracy of 97.99 $\pm$ 1.96\% across all participants at classifying between the four different vergence gestures. From the per-class perspective presented in Fig.~\ref{fig:confusion_matrix}(a), our system achieved near-perfect discrimination as shown in the confusion matrix.}

\modify{We also evaluate our system on all six gestures supported by the three selected distances. Our evaluation (Table~\ref{tab:statistical_res}) shows that the system achieves a within-session classification accuracy of 82.68 $\pm$ 4.29\%.} There is virtually no misclassification between the 30 $\rightarrow$ 200~cm and 70 $\rightarrow$ 200~cm transitions, or between the 200 $\rightarrow$ 70~cm and 200 $\rightarrow$ 30~cm transitions, which is similar to the results for four gestures. However, the 30 $\rightarrow$ 70~cm movement creates an EOG waveform that can overlap with the 30 $\rightarrow$ 200~cm and 70 $\rightarrow$ 200~cm movements, and the 70 $\rightarrow$ 30~cm movement likewise can overlap between the 200 $\rightarrow$ 70~cm and 200 $\rightarrow$ 30~cm movements. 

\add{Importantly, for both the 4- and 6-gesture evaluations the near-to-far and far-to-near vergence movements are never confused, consistent with the observation in Sec.~\ref{sec:distance_selection}.}


\noindent {\bf Cross-Session Performance.} When the user remounts the glasses, slight shifts in electrode placement introduce signal drift. To evaluate the robustness of our system against remounting, we trained a model on one of the user’s data collection sessions and tested it on the other unseen session. We then reversed the training and testing sessions and averaged the results. Our evaluation (Fig. ~\ref{fig:per-user-comparison}b, Fig.~\ref{fig:confusion_matrix}b, Table~\ref{tab:statistical_res}) \modify{shows that system performance remained stable at 97.78\% for the four-gesture task, validating the robustness of our system against signal drift induced by glasses remounting. However, for the 6-gesture detection task, the system performance decreased from 82.68\% to 77.78\% after remounting.}


\noindent {\bf Cross-User Performance.} We evaluate the performance of a pretrained model in a leave-one-user-out setup across the 20 participants in our study. Results in Table~\ref{tab:statistical_res} show that our system achieves an accuracy of \modify{97.01\% and 77.36\% without any new-user enrollment for the 4-gestures and 6-gestures tasks respectively}.

\begin{table}
\centering
\begin{tabular}{|c|c|c|c|}
\hline
\textbf{} & \textbf{Within Session} & \textbf{Cross Session} & \textbf{Cross User} \\
\hline
\makecell{4-gesture\\accuracy} & 97.99 $\pm$ 1.96\% & 97.78 $\pm$ 2.02\% & 97.01 $\pm$ 2.06\% \\ \hline
\makecell{6-gesture\\accuracy} & 82.68 $\pm$ 4.29\% & 77.78 $\pm$ 6.17\% & 77.36 $\pm$ 6.17\% \\ \hline
\end{tabular}
\caption{Classification performance (mean $\pm$ stdev) for \modify{4- and 6-gesture} classifier. \add{The primary use case for our system targets a 4-gesture classifier.}}
\label{tab:statistical_res}
\vspace{-2em}
\end{table}

\noindent {\bf Effect of Vertical Channels.}  
We assessed the effect of vertically placed electrodes on system performance across 9 participants. As shown in Fig.~\ref{fig:per-user-comparison}b, 5-fold cross-validation shows that our two-channel configuration \add{achieved 98.61\% accuracy across four gestures, while the three- and four-channel configurations result in modest performance gains of 0.28\% and 0.42\% and accuracies of 98.89\% and 99.03\% respectively.} These modest gains are likely because vergence is primarily, though not exclusively, a horizontal eye movement. However, all participants reported discomfort with the 3- and 4-channel configurations during data collection. Given these persistent ergonomic concerns, we excluded these configurations from the protocol for the remaining 11 participants. These findings further justify our decision to prioritize a two-channel design for the final system.

\subsection{Subgroup analysis}
\label{sec:subgroup_analysis}
\begin{figure*}
\includegraphics[width=0.9\linewidth]{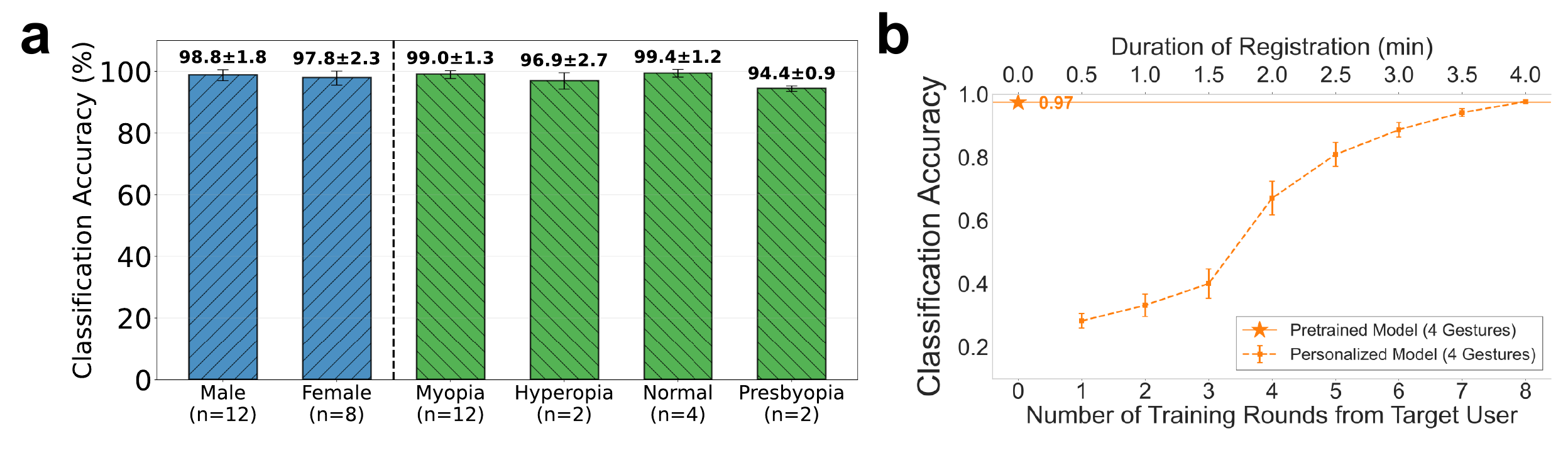}
    \caption{\textbf{(a)} Subgroup analysis on within-session classification accuracy for 4-gestures. {\bf (b)} Effect of training duration on classification accuracy.}
    \label{fig:subgroup}
\end{figure*}

To evaluate the generalization and robustness of {\sysname} across diverse user groups, we conducted a subgroup analysis on the within-session data across 2 sessions. We assessed classification performance across demographics and physiological conditions, as shown in Fig.~\ref{fig:subgroup}a.

\noindent \textbf{Sex.} When comparing the classification performance by sex, the female participants ($n=8$) achieved a mean accuracy of \modify{97.81 $\pm$ 2.29\% for 4-gesture task and 81.56 $\pm$ 4.67\% for 6-gesture task.} The male participants ($n=12$) achieved \modify{98.75 $\pm$ 1.77\% and 83.54 $\pm$ 3.98\% for 4-gesture and 6-gesture tasks}, respectively.

\noindent \textbf{Vision conditions.} Participants who normally wear corrective lenses were instructed to wear them with our {\sysname} glasses on top of them during the study to ensure they could clearly focus on vergence targets at 30, 70 and 200~cm. The mean classification accuracies for the tasks of \modify{4-class and 6-class, respectively, were as follows: 98.96 $\pm$ 1.29\% and 84.03 $\pm$ 4.92\% for myopia ($n=12$); 99.38 $\pm$ 1.25\% and 82.91 $\pm$ 1.98\% for normal vision ($n=4$); 96.88 $\pm$ 2.65\% and 79.58 $\pm$ 0.60\% for hyperopia ($n=2$); and 94.38 $\pm$ 0.88\% and 77.90 $\pm$ 1.77\% for presbyopia ($n=2$).}


    

    
\subsection{\modify{Comparison against baseline electrode layouts}} 
\label{sec:validate_electrode}
We perform a comparison against two baseline EOG sensing systems.
First, we compare our system with Mishra et al.’s work~\cite{mishra2020soft}, which is focused on vergence sensing using custom aerosol-jet–printed, stretchable, skin-like biopotential electrodes using a 2-, 3-, and 4-channel configurations. Their 2-channel configuration differs from ours and has a horizontal channel across both eyes, a vertical channel above and below the right eye, and a ground electrode obtrusively placed on the forehead, and reports an accuracy of 34\%. In contrast, our system uses an unobtrusive 2-channel configuration using standard Ag/AgCl electrodes with the ground on the mastoid and achieves an accuracy of 82.68\%. While their 3-channel and 4-channel configurations have increased accuracies of 87 and 91\% respectively, additional electrodes bring about discomfort, which negatively affects the user experience.

\begin{table}
\centering
\begin{tabular}{|l|c|c|}
\hline
\textbf{} & \textbf{Mishra et al.~\cite{mishra2020soft}} & \textbf{{\sysname} (ours)} \\
\hline
Number of electrodes & 5 - 7 & 4\\ \hline
Number of channels & 2 - 3 & 2\\ \hline
In-session accuracy & 34\% - 91\% & 82.68\% \\ \hline
Cross-session accuracy & Not specified & 77.78\% \\ \hline
Cross-user accuracy & Not specified & 77.36\% \\ \hline
Commodity electrodes & \xmark & \cmark\\ \hline
Motion artifact removal & \xmark & \cmark\\ \hline
Ground electrode & \makecell{Forehead\\(obtrusive)} & \makecell{Mastoid\\(unobtrusive)} \\ \hline
Glasses Compatibility & \xmark & \cmark \\ \hline
Angular difference tested & 1--3 & 1.35--4.05 \\ 
\hline
\end{tabular}
\caption{Comparison between our system and the baseline proposed by Mishra et al. ~\cite{mishra2020soft}}
\label{tab:comparison_science_advanced}
\vspace{-2em}
\end{table}

Second, we compare our system's performance against the EOG electrode layout in the commercial JINS MEME glasses~\cite{jinsmeme} (Fig.~\ref{fig:electrode_placement}) \add{and left-right differential layout represented by E-Glasses~\cite{Lee20203DPrinted}.} To ensure a fair comparison, one participant (P4) wore a combined electrode setup encompassing the positions of both our system and the JINS MEME. We then collected data across ten rounds, each including all six vergence gestures. \modify{Evaluating each configuration independently using five-fold cross-validation, our system achieves an accuracy of 100\% in 4-gesture task, outperforming the JINS MEME layout (67.5 $\pm$ 15\%) \add{and the E-Glasses channel (72.5 $\pm$ 9.35\%)}. In the more complex 6-gesture task, our system achieved 88.33 $\pm$ 1.13\% accuracy, whereas JINS MEME and the derived E-Glasses channel achieved only 50.0 $\pm$ 1.29\% and 53.33 $\pm$ 8.50\%, respectively. These results are in line with our prior SNR analysis, which demonstrated that our vergence waveforms exhibit an average SNR improvement of 7.3~dB over JINS MEME and 4.8~dB over the E-Glasses configuration.}


\subsection{\modify{Comparison of classifier models}}
\label{sec:eval_baseline}



\orange{We compared our feature-based Random Forest pipeline with 1D-CNN and LSTM models operating directly on filtered two-channel EOG segments. The model architectures and tuning details are provided in Appendix~\ref{sec:appendix_baseline}. As shown in Table~\ref{tab:classification_baselines}, Random Forest achieved the highest within-session (82.68 $\pm$ 4.29\%) and cross-session accuracy (77.78 $\pm$ 6.17\%), suggesting that the handcrafted features effectively represent the distinguishable patterns of gesture. In the cross-user setting, the random forest and 1D-CNN had comparable results.}

\begin{table}[H]
\centering
\small
\begin{tabular}{lccc}
\hline
\textbf{Model} & \textbf{Within-session} & \textbf{Cross-session} & \textbf{Cross-user} \\
\hline
Random Forest & \textbf{82.68 $\pm$ 4.29\%} & \textbf{77.78 $\pm$ 6.17\%} & 77.36 $\pm$ 6.17\% \\
1D-CNN & 77.13 $\pm$ 5.89\% & 73.56 $\pm$ 6.90\% & \textbf{77.64 $\pm$ 4.82\%} \\
LSTM & 74.63 $\pm$ 6.66\% & 65.79 $\pm$ 7.51\% & 73.66 $\pm$ 5.96\% \\
\hline
\end{tabular}
\caption{\add{\textbf{Comparison of different classifiers for 6 vergence gestures.}}}
\label{tab:classification_baselines}
\vspace{-1em}
\end{table}

\section{\orange{Evaluations toward real-world deployment}}



\subsection{\modify{Compatibility with non-vergence gestures}}
\label{sec:11way}

\begin{figure}
    \centering
    \includegraphics[width=.8\linewidth]{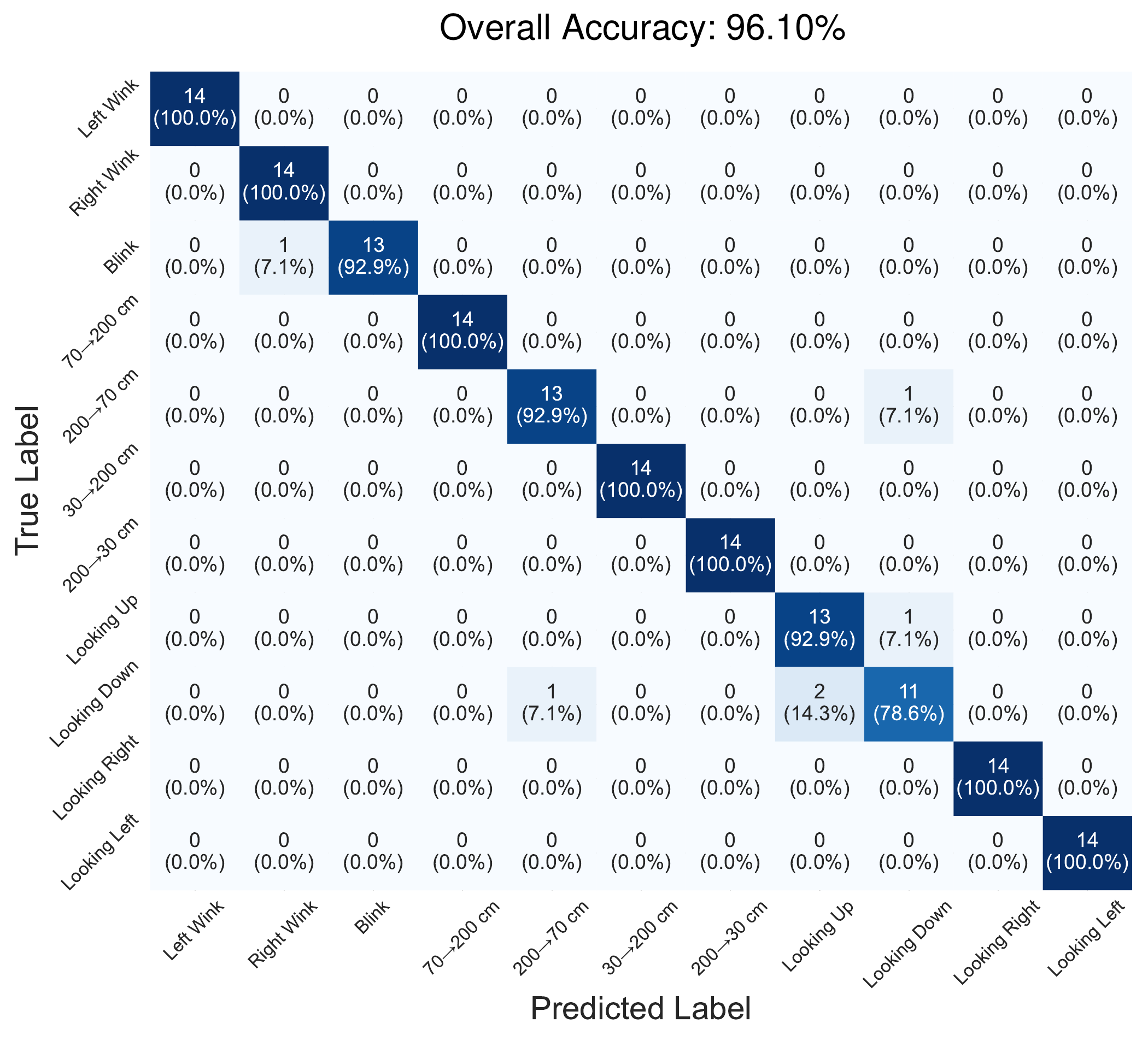}
    \caption{Confusion matrix classifying between different vergence and non-vergence movements across 11 classes.}
    \label{fig:11way}
    \vspace{-1em}
\end{figure}

\modify{Prior smart-glasses EOG interfaces typically support a vocabulary of conventional eye movements, such as blinks, winks, and directional gazes. We therefore evaluate whether our system preserves this compatibility while enabling vergence sensing.} We collect a dataset including the four vergence gestures with high separability, along with seven additional eye movements: left and right winks, blinks, and four directional gazes (up, down, left, right). We collected 14 recordings for each class, and performed 5-fold cross-validation. As shown in Fig.~\ref{fig:11way}, our system achieved an accuracy of 96.1 $\pm$ 2.4\%.

A few misclassifications were observed between upward and downward gaze, which is expected as our design intentionally omits vertical electrode channels to avoid user discomfort. However, our system still captures some vertical eye movements as we intentionally incorporate vertical separation between positive electrodes and reference electrode, as shown in Fig.~\ref{fig:electrode_placement}c. These results show that our electrode layout \orange{can effectively detect eye movements supported by prior smart-glasses EOG systems alongside vergence gestures, expanding the interaction vocabulary available on smart glasses.}



\subsection{Generalizing to virtual depth cues}

\modify{Beyond physical Brock String targets, a practical smart-glasses interface must operate with rendered content. We therefore evaluate whether {\sysname} can detect vergence induced by stereoscopic disparity. This simulates AR/VR applications where users focus on virtual objects rendered at different depths on a fixed-depth display.} 

\begin{figure}
    \centering
    \includegraphics[width=1\linewidth]{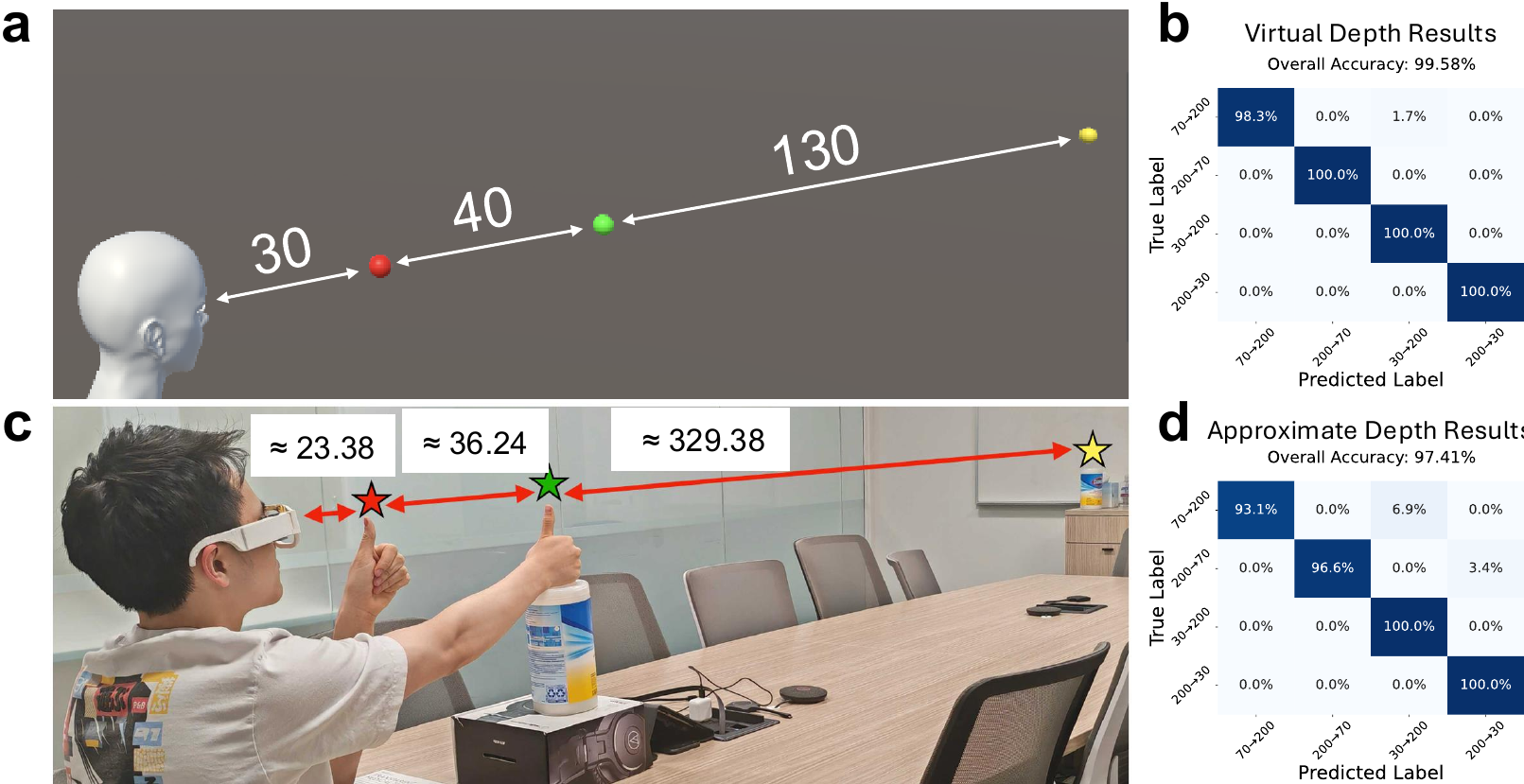}
    \caption{Experimental setup to evaluate generalization to {\bf (a)} virtual depth cues and {\bf (c)} approximate depth cues. Results are shown as confusion matrices \modify{across five participants for the four-gesture task} as {\bf (b)} and {\bf (d)} respectively. \add{Distances are in cm.}}
    \label{fig:thumb}
    \vspace{-1em}
\end{figure}

To test this, we developed a prototype application on the Meta Quest 2~\cite{metaquest2}, replicating the Brock string setup described in Sec.~\ref{sec:user_study_design} within a Unity-based VR environment. Beads were virtually placed at 30, 70, and 200~cm from the user’s eye center (Fig.~\ref{fig:thumb}a). \modify{We then invited five participants to perform the same vergence gestures following the procedure described in Sec.~\ref{sec:user_study_design}, yielding 300 virtual-depth test gestures in total.}

We trained our model only on the Brock String dataset, while excluding Brock String sessions from the same target participant when applicable. {\sysname} achieved an accuracy of 99.58 $\pm$ 0.77\% and 84.00 $\pm$ 4.42\% for four- and six-gesture classification respectively (Fig.~\ref{fig:thumb}b). This result shows that {\bf (1)} virtual depth overlays can reliably induce vergence, even though they are rendered on a fixed display, and {\bf (2)} a model trained on \orange{physical Brock String} depth cues can successfully generalize to virtual depth cues.

\subsection{Generalizing to approximate depth cues} 
\label{sec:approximate_depth}

\modify{We next evaluate if our system can work in scenarios where precise physical targets are unavailable. To do this, we propose a body-referenced interaction method where the user fully extends one arm, rests the other hand at the elbow of the extended arm, and aligns both thumbs with a far-field object. This setup approximates the distances of 30 and 70~cm used in the Brock string experimental setup, and corresponds to a half and full arm’s length for an average adult~\cite{roberts1959arm}. While the Brock string setup used 200~cm as the far-field depth cue, in this scenario the exact depth of the far-field cue is less critical, as vergence-induced EOG signal changes diminish with increasing distance (Fig.~\ref{fig:depth-analysis}).}

\add{We evaluate generalizability for four gestures as the primary use case, where the user alternates focus between the thumbs and a far-field object regarding far-field target as an anchor. For richer input, this can be extended to the six gesture case using both thumbs and a far-field object as reference points. We had five participants perform the vergence gestures following the protocol in Sec.~\ref{sec:user_study_design}. We note that due to differences in arm length across participants, this led to a natural variation in the depth of the near and middle cue, which averaged 23.38 $\pm$ 5.03~cm (ranging from 15.6 to 29.1~cm) and 59.62 $\pm$ 6.48~cm (ranging from 51.5 to 68.0~cm), respectively. We used objects at diverse distances as far cues and were at distances averaging 389.00 $\pm$ 110.43~cm (ranging from 255.0 to 492.0~cm). We trained our model only on the Brock String dataset, while excluding Brock String sessions from the same target participant when applicable.}

\modify{{\sysname} achieved 84.00 $\pm$ 4.42\% accuracy on the six-class task. For the four-gesture task, the model achieved 97.41 $\pm$ 4.04\% accuracy over the 10 rounds data (Fig.~\ref{fig:thumb}d). When collapsing predictions into a binary direction-level task: near-to-far vs. far-to-near, the model achieved 100\% accuracy, indicating that opposite-direction depth transitions were not confused. These results demonstrate that our system generalizes to approximate physical depth interactions without requiring the physical cues to exactly match the original Brock String distances.}

\subsection{Motion artifact and noise removal performance}
\label{sec:motion_eval}

\add{While previous evaluations focused on vergence detection, real-world deployment poses a challenge in rejecting artifacts caused by facial, eye, and body movements. We therefore evaluate whether our artifact-removal pipeline can \orange{identify and reject} these non-vergence events while reliably preserving vergence inputs.}

We evaluated the motion artifact and noise removal pipeline on a \modify{dataset collected from $n = 6$ participants (Table~\ref{tab:recordings}). The evaluation employs a 2-second moving-window pipeline. We split training and test sets since neighboring windows are highly correlated. This prevents data leakage by ensuring that overlapping windows from the same recording never appear in both sets.}

\add{We used a personalized artifact-removal model where participants would perform 6 minutes of calibration: 3 minutes for six rounds of vergence data and 3 minutes for one artifact round for each of the six motion/noise activities. This resulted in an accuracy of 98.77 $\pm$ 1.26\% across six participants, with a false positive rate of 1.23 $\pm$ 1.26\%. Detailed calibration-budget ablations and cross-user adaptation results are provided in Appendix~\ref{sec:appendix_motion_artifacts}.}

\begin{figure*}
\includegraphics[width=.5\linewidth]{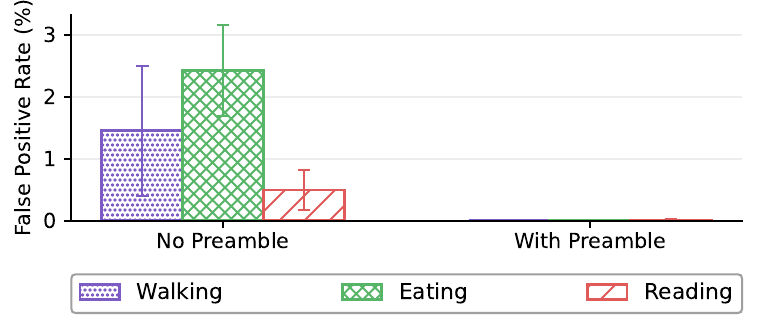}
    \caption{False positive rates \modify{across everyday activities: eating, walking, and reading.} A preamble consisting of an eyebrow raise to activate the system reduces the false positives to \modify{nearly zero across all three} conditions.}
    \vspace{-1em}
    \label{fig:curve}
\end{figure*}

\subsection{False positives from non-vergence activity} 
\label{sec:false_positives}

\modify{We further conducted an online false-positive evaluation during non-vergence sessions where participants wore the prototype while performing daily activities. The artifact-removal pipeline ran in real-time on the prototype during the evaluation. Six participants completed three tasks: eating while using a phone or watching a movie (10.52 $\pm$ 0.41 minutes), walking while conversing (11.12 $\pm$ 0.45 minutes), and reading a document on a computer screen (10.83 $\pm$ 0.53 minutes). To broaden evaluation environments, eating and reading were conducted in separate indoor rooms, while walking was conducted outdoors.}

\modify{In this evaluation, a false positive is defined as a non-vergence sliding window falsely accepted by the system. As shown in Fig.~\ref{fig:curve}, using the frozen classifier trained on the data described in Sec.~\ref{sec:motion_eval}, we obtained a false positive rate of 2.43 $\pm$ 0.73\% during eating. Walking while conversing and reading on a computer produced lower false positive rates of 1.45 $\pm$ 1.05\% and 0.50 $\pm$ 0.32\%, respectively.} To further reduce the false positive rates, we introduced a brow-raise preamble gesture to explicitly activate and deactivate the system. The motion is distinct from other movements, and results in an EOG amplitude that is significantly higher than vergence (Fig.~\ref{fig:segments}), and it is easy and quick to perform. \modify{After incorporating this preamble, the false positive rate dropped to nearly zero across all three activities: 0.005\% $\pm$ 0.005\% for eating, 0\% for walking, and 0.015\% $\pm$ 0.015\% for reading.}

\subsection{Power consumption}
\label{sec:power_consumption}


We measured the power consumption of our system using the Monsoon High Voltage Power Monitor~\cite{monsoon} at 3.7~V, over 5 runs and averaged them, with each run lasting 1 minute at the lowest clock speed of 24~MHz.

The power consumption of the two amplifier front-ends was 3.3 $\pm$ 0.004~mW. This is lower than other eye tracking systems, in particular mmWave radios~\cite{ma2025mmet}, and microphones and speakers~\cite{li2024gazetrak}, though we note that our system is purpose-built for the specific task of vergence gesture sensing not eye-tracking.

The code for data acquisition (digitization, reading samples into buffers) consumed  4.84 $\pm$ 0.02~mW; while data processing (filtering, segmenting, classification) consumed 2.63 $\pm$ 0.02~mW.

With a 570~mAh rechargeable lithium-ion battery as in Google Glass~\cite{gglass_battery}, the system would operate for approximately 197 hours (8.2 days). This power draw is effectively negligible, as the overall battery life of Google Glass is dominated by other processes that result in a typical battery life of 8 hours~\cite{gglass}. \add{Further, we note that a productized version of the system would consolidate the sensing, data acquisition and processing into a low-power ASIC or SoC. However, the design of an integrated system is out of scope for this work.}

We note that other standby processes on the microcontroller consume on average 138.2 $\pm$ 0.0741~mW. However, we do not consider this as part of the power consumption of our proof-of-concept system, as a commercial EOG-based system would leverage a low-power ASIC or SoC, which we leave for future work.

\subsection{System latency} 
\add{We evaluate whether the processing pipeline can support real-time interaction by measuring} the wall-clock time of our data processing pipeline, which continuously filters and classifies the streaming EOG signal 2-second rolling windows with a 0.1-second overlap. The average latency across 2500 windows was 15 $\pm$ 0.002~ms, which is less than the step duration of 0.1 seconds and demonstrates our system's ability to run in real-time.

\subsection{User Experience Survey}
\label{sec:user_experience}

\add{Finally, we assess subjective aspects of the user experience including electrode comfort, facial-movement interference, perceived fatigue, accidental-trigger concern, and social acceptability.} We evaluated the user experience during the experiment with a survey. We asked the following questions: 

\begin{enumerate}
    \item How easy is it to keep the electrodes on? (1 = Hard, 5 = Very easy to keep on)
    \item How easy was it to move your facial muscles (e.g. raising your eyebrows, closing your eyelids)? (1 = Very difficult, 5 = Not Difficult)
    \item How concerned are you about accidentally performing these eye gestures in your daily life? (1 = Very concerned, 5 = Not at all)
    \item To what extent did you feel physical or mental fatigue when performing an eye gesture with the system? (1 = High fatigue, 5 = No fatigue)
    \item How comfortable would you feel performing these eye gestures in a public setting? (1 = Very uncomfortable, 5 = Very comfortable)
    \item How comfortable would you feel performing these eye gestures in a private setting? (1 = Very uncomfortable, 5 = Very comfortable)
\end{enumerate}


The survey results are summarized in Fig.~\ref{fig:survey}. Participants ($n=20$) had an average age of $25.75 \pm 11.1$ years and height of $171.83 \pm 7.65$~cm. \orange{Participants reported that the electrodes remained securely in place (3.85 $\pm$ 0.93) and that facial movements remained easy to perform (4.10 $\pm$ 0.98). Comfort was higher in private settings (4.75 $\pm$ 0.55) than in public settings (3.75 $\pm$ 0.91), while fatigue rating was 3.55 $\pm$ 0.89. Additionally, users also reported confidence in in performing the movement intentionally on command, with the concern about accidentally performing the gesture in daily life received an average score of $3.75 \pm 1.25$.}

\begin{figure*}
\includegraphics[width=1.0\linewidth]{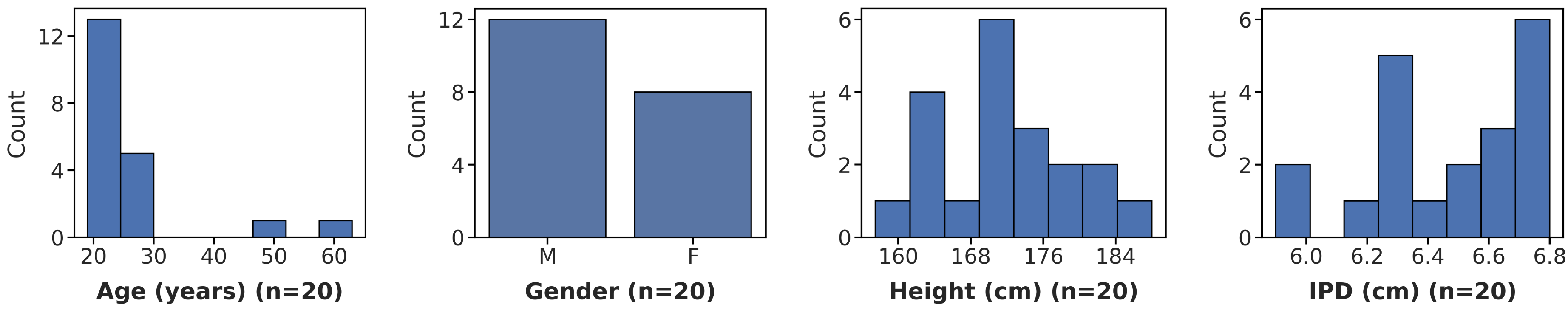}
    \caption{{\bf Demographic summary of participants in the user study.} Survey data was collected on $n=20$ participants in the data collection procedure.}
    \label{fig:demographic}
\end{figure*}

\begin{figure*}
\includegraphics[width=1.0\linewidth]{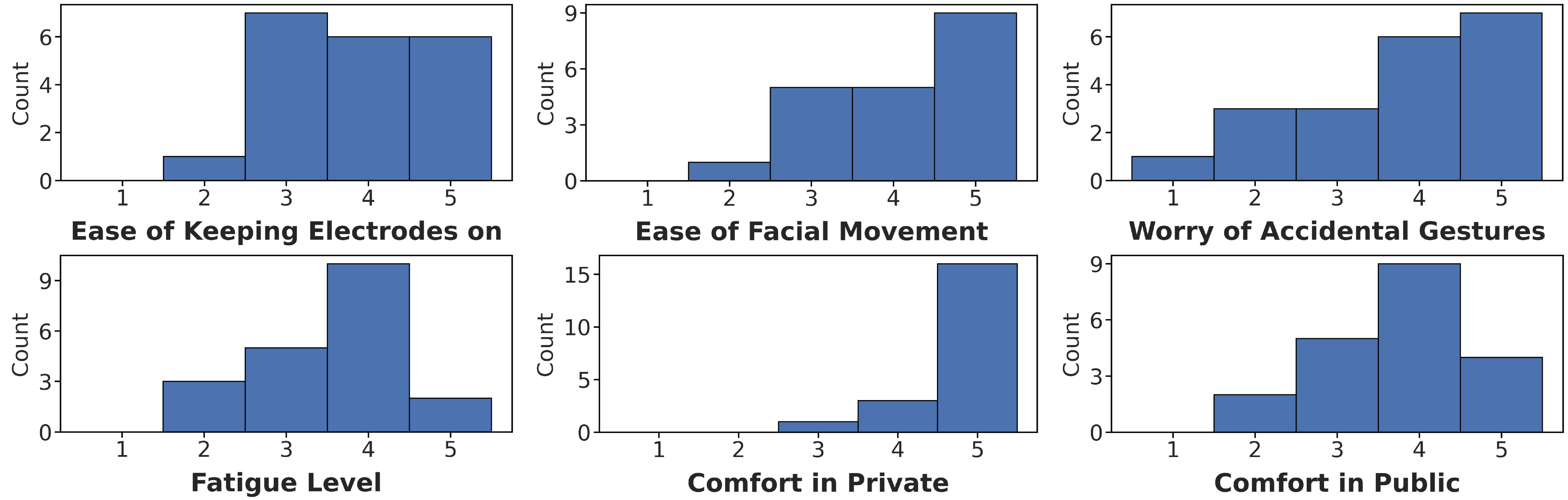}
    \caption{{\bf User experience survey ($n = 20$).} Histograms summarize user perceptions across (a) ease of maintaining electrode contact, (b) ease of facial muscle movement while wearing the prototype, (c) concern regarding accidental gesture triggers, (d) interaction-induced fatigue, and the social acceptability and comfort of the interaction in (e) private and (f) public settings.}
    \label{fig:survey}
\end{figure*}

\section{Example applications}
\label{sec:applications}


\subsection{Vergence-driven varifocal glasses.}
\label{sec:application_varifocal}
%

\orange{Varifocal glasses adjust optical focus to match the user’s viewing distance~\cite{padmanaban2018autofocals}, helping billions of people with impaired accommodation, including presbyopia, focus across near and far distances~\cite{holden2008global}. Commercial efforts including IXI~\cite{ixi}, ViXion01~\cite{ViXion}, and Elcyo~\cite{elcyo} demonstrate growing interest in controlling varifocal glasses automatically, but these systems rely on camera-based sensing, which can be power-intensive for continuous wear. }In contrast, we present the first attempt to leverage EOG signals as a low-power, lightweight modality for \add{future} automatic varifocal control. Importantly, although the eye lens' focusing ability is impaired, individuals can still perform vergence movements betweeen near and far distances ~\cite{padmanaban2018autofocals}. We have also validated this in Sec.~\ref{sec:subgroup_analysis}, where our system achieved $94.38 \pm 0.62\%$ 4-gesture accuracy for individuals with presbyopia. 

\begin{figure*}
    \centering
    \includegraphics[width=1\linewidth]{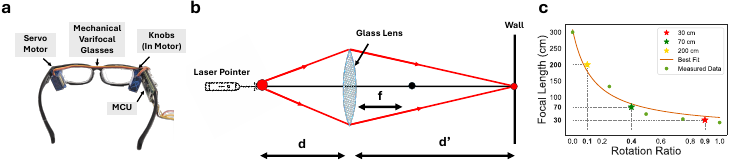}
    \caption{{\bf (a)} Prototype of mechanically adjustable varifocal glasses with an integrated servo motor for actuation. {\bf (b)} Experimental setup to determine the effective focal length at each varifocal lens setting. {\bf (c)} Effective focal length of the varifocal glasses for different knob rotation ratios.}
    \label{fig:varifocalglass}
\end{figure*}

Our prototype (Fig.~\ref{fig:varifocalglass}a) instruments an existing pair of varifocal glasses~\cite{varifocal} with a small servo motor~\cite{servo} (\( 22.8\,\mathrm{mm} \times 12.2\,\mathrm{mm} \times 28.5\,\mathrm{mm} \)) which is coupled to a knob to adjust the focal length. The motor weighs 29~g, which is relatively modest compared to commercial eye-tracking glasses such as the Tobii Pro Glasses 3 (388~g)~\cite{tobii}.

To show that our adjustable varifocal glasses can support focusing at distances of 30, 70, and 200\,cm, we conducted a laser-based experiment to measure the effective focal lengths of the lens. The goal was to determine the focal length corresponding to different rotations of the glasses' adjustment knob. In the setup (Fig.~\ref{fig:varifocalglass}b), a laser pointer was aimed at a distant wall with the varifocal lens placed in between. Let $d$ be the distance from the laser to the lens and $d'$ the distance from the lens to the wall. We adjusted $d'$ by moving the varifocal glasses until the laser spot appeared in focus, then computed the effective focal length $f$ using the lens equation~\cite{lens_eq}: $f = \frac{d \cdot d'}{d + d'}$. The total distance between the laser and the wall was 12~m, which allowed us to measure focal lengths up to 3~m. 

This procedure was repeated for five normalized knob rotations ranging from 0.0 to 1.0. As shown in Fig.~\ref{fig:varifocalglass}c, the effective focal length spans approximately \( 25.58\,\text{cm} \) to over \( 300\,\text{cm} \), covering the key distances required for near-to-far visual accommodation in individuals with presbyopia.

\subsection{Stereoscopic depth rendering of virtual objects.}


Modern AR displays often render virtual objects at varying depths using stereoscopic rendering, which displays slightly different images to each eye to create a sense of depth from binocular disparity~\cite{blonde20103d,wan2004interactive}. Vergence sensing can be used to align virtual content with the user’s depth of focus. Importantly, although all virtual content is rendered on a screen located at a fixed physical distance, stereoscopic rendering can still induce vergence.
\begin{figure}
    \centering
    \includegraphics[width=.6\linewidth]{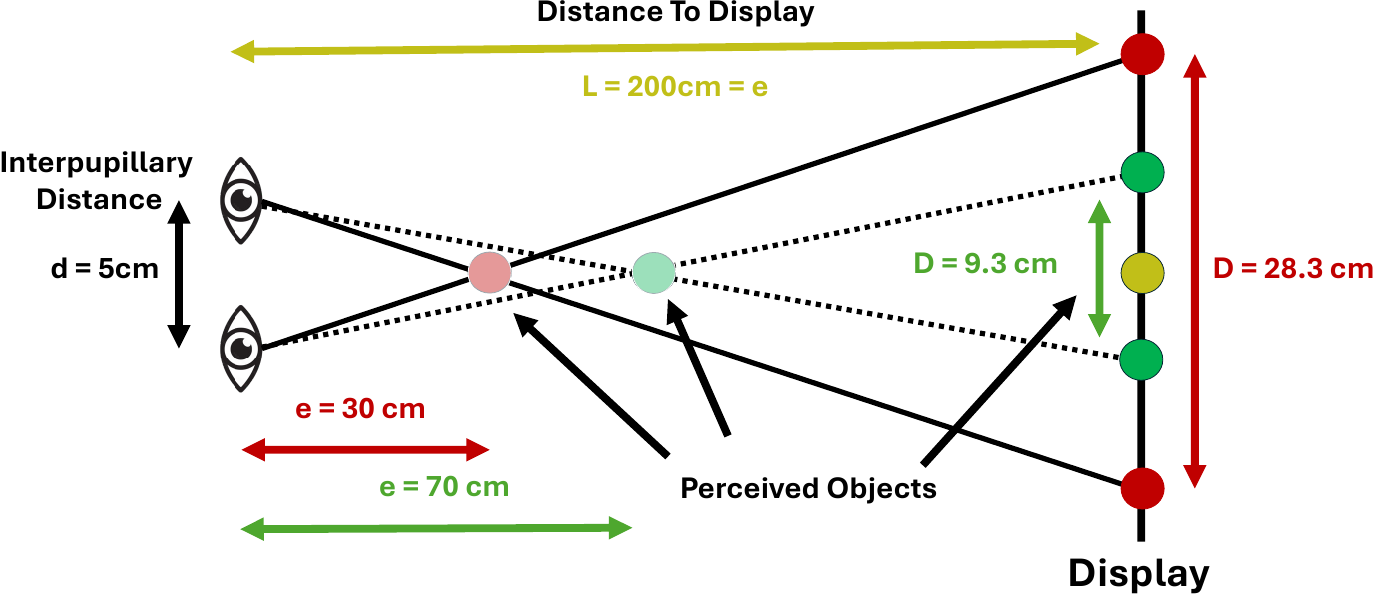}
    \caption{Stereoscopic depth rendering is achieved by offsetting the left and right images to simulate the three depths (30, 70, 200~cm) supported by our system.}
    \label{fig:stereo}
    \vspace{-1em}
\end{figure}
Fig.~\ref{fig:stereo} illustrates the stereoscopic depth estimation formula~\cite{John1941Stereo}: $D = d \cdot \left( \frac{L}{e} - 1 \right)$ which computes the required separation $D$ between the left and right projected images to make an object appear at a desired virtual depth $e$. Here, $L$ is the distance between the eyes and the screen, and $d$ is the interpupillary distance. 

In the example, the user is positioned $L=200~cm$ from the screen, with an interpupillary distance of $d=5~cm$. To render an object at $e=30~cm$ and $e=70~cm$, the images would need to be separated by $D=28.3~cm$ and $D=9.3~cm$ respectively. While to render it at $e=200~cm$, which matches the physical distance to the screen, the images should overlap, resulting in zero separation.

\section{Discussion, limitations, and future work}
\noindent {{\bf \orange{Interaction value of vergence.}}} \orange{Existing EOG glasses can detect blinks, winks, dwell-based gaze, and directional gaze for generic commands or 2D angular selection. In contrast, vergence offers two complementary interaction benefits. First, it provides a natural mapping between a shift in gaze depth and an action associated with the corresponding depth cue. Furthermore, vergence can also serve as an intentional cue that has been leveraged to mitigate the ``Midas touch'' problem~\cite{ahn2020verge,zhang2024focusflow}, in which natural eye movements detectable by prior EOG systems are mistaken for deliberate commands~\cite{Jacob1990Midas}.}

\noindent {{\bf Learnability of vergence.} 
Vergence is a learnable and controllable skill even without visual reference points, enabling users to perform intentional depth-based eye gestures with training in a few minutes~\cite{mclin1988voluntary,hamori1982study,zhang2024focusflow,kirst2016verge,kudo2013input,ruan2018human}. An example voluntary control is perceiving a 3D autostereogram~\cite{tyler1990autostereogram} using cross-eyed or wall-eyed viewing.} 

\noindent {\bf Practical vocabulary size for vergence gestures.} Evaluation results show that our system reliably supports 4-gesture vergence interaction, achieving over 97\% accuracy both after remounting and when generalizing to new users without calibration. In contrast, our system is less reliable when detecting all six gestures across near, mid, and far distances. Most errors come from the 30 $\leftrightarrow$ 70~cm transitions, as the signals overlap with other gestures. \orange{This issue may come from both sensor limitations and variability in the oculomotor response.} From a hardware perspective, prior work shows that even with high-quality wet electrodes, tracking both eyes has an uncertainty of about 1.5° for each eye~\cite{Young1975}, resulting in a 3° uncertainty. In our setup, the 2.7° binocular angular difference is inherently smaller than this theoretical threshold, making it difficult to separate reliably using dry electrodes. \orange{From a physiological perspective, the human vergence response consists of a fast open-loop initiation followed by a slower, visually guided closed-loop feedback phase~\cite{semmlow2019dynamics}. The resulting inter-subject variability in execution speed and trajectory makes gesture boundaries even harder to distinguish. In summary, fine-grained vergence sensing on EOG glasses remains challenging and we recommend using four-gesture as the default interaction vocabulary.}




\noindent \textbf{\orange{Vergence with gaze direction changes.}} \orange{Prior work shows that simultaneous changes in gaze direction and depth can alter vergence dynamics~\cite{kumar2005directional}. Such changes were unavoidable in our body-referenced study in Sec.~\ref{sec:approximate_depth}, yet our system achieved 97.41\% accuracy. Future work should systematically examine the joint effects of gaze direction and depth with the help of high-precision eye trackers.}

\noindent \textbf{\add{Depth granularity.}} \add{Our choice of depth distances is grounded in the practical system design discussed in Sec.~\ref{sec:distance_selection}. The selected near, mid, and far distances not only cover common viewing ranges but also align with the focal zones used in optical systems. However, our current study does not explore distances at a finer granularity. For example, we have not investigated the minimum angular difference required to distinguish two gestures. It also remains unclear whether the same angular change produces similar EOG responses at near and far viewing distances. A deeper study of depth would require more distance levels, finer spacing between targets, and synchronized ground-truth eye tracking to analyze how vergence angle, movement dynamics, and EOG responses vary across various distances. We view this as an important direction for future work.}

\noindent \add{{\bf Adapting to different anatomies.} A fundamental challenge for EOG-based eye movement sensing systems is balancing robust electrode-skin contact with user comfort. Our design addresses this balance through optimized structural features including a flexible nose-bridge and curved mastoid support. Future work could also leverage signal quality and user feedback to dynamically guide device positioning or trigger automated micro-adjustments.}



\noindent \textbf{Preamble mechanism to mitigate accidental triggers.} To \orange{support} system reliability against daily facial activities (e.g. chewing, speaking), our current design incorporates raising eyebrows as an explicit preamble for activation. Although this mechanism effectively mitigates false positives, requiring a deliberate physical gesture may be intrusive in social contexts and interrupt natural interaction flows. We aim to evolve towards implicit activation mechanisms in future iterations, potentially by fusing EOG with lightweight IMU data or leveraging context-aware algorithms to seamlessly discern interaction intent without conscious user effort.

\noindent \textbf{\orange{New user enrollment.}} \orange{Although {\sysname} can distinguish four vergence gestures on unseen users without calibration, we found that artifact rejection remains user-dependent as electrode-skin contact, facial anatomy, and motion patterns differ across users. {\sysname} therefore requires a new user to perform six repetitions of each gesture and approximately three minutes of motion/noise activities to personalize artifact rejection.}

\noindent {\bf Long-term usage.} Our evaluation was conducted primarily in controlled settings. For future work, longitudinal in-the-wild studies are needed to assess system performance in the presence of perspiration, electrode displacement, skin-impedance drift, \orange{spontaneous facial movements, as well as user comfort over hours or days.}

\section{Conclusion}
This paper presented {\sysname}, the first EOG-based glasses system that enables robust, depth-aware eye interaction by sensing vergence. With a \orange{glasses-compatible} electrode layout with positive electrodes on the temples and a shared negative electrode on the nose bridge, \orange{{\sysname} preserves the opposing-polarity signals from the two eyes and captures high-SNR vergence responses.} \modify{Our evaluation across 20 users demonstrated that {\sysname} can distinguish four vergence gestures on unseen users with 97.01\% accuracy without calibration.} With a front-end power consumption of only 3~mW and a processing latency of 15~ms, the system \orange{supports low-power, real-time vergence sensing.} We envision that the sensing capability introduced by {\sysname} will broaden the scope of eye interaction available on EOG-based glasses, serving as a lightweight and privacy-preserving primitive that enables a new generation of depth-aware applications in consumer electronics, healthcare, and safety-monitoring domains.

\section{LLM Use Clarification}
Gemini-3 Pro was used for generating human characters in Fig. \ref{fig:timeline}, as well as for grammar correction and text refinement throughout this paper.


\newpage
\appendix
\section{\add{Feature Selection for Vergence Classification}}
\label{sec:appendix_feature}

\add{Prior EOG-based vergence work has shown that compact time-domain features can effectively represent vergence-related EOG windows for real-time classification. For example, Mishra et al.~\cite{mishra2020soft} used handcrafted features such as signal integral, amplitude, velocity, and variance, further evaluating candidate features through wrapper-based selection. Motivated by prior work, we designed a compact feature representation tailored to the morphology of the two-channel vergence signals captured by our glasses.}

\add{As illustrated in Fig.~\ref{fig:depth-analysis}c, vergence EOG segments typically contain a dominant extremum: transitions from near to far produce a valley in both EOG channels, whereas transitions from far to near produce a peak. We therefore split each detected vergence segment at its dominant extremum. This produces two temporal portions corresponding to the approach-to-extremum and recovery phases of the vergence response.}

\subsection{\add{Proposed Compact Feature Representation}}

\add{For each half-segment, we extract five temporal features: amplitude range, area under the curve, end-to-end slope, mean of the first derivative, and variance of the first derivative. These descriptors capture the magnitude of the response, the strength of the accumulated signal, the coarse temporal trend, the average local change and the local temporal variability. This yields 10 features per EOG channel.}

\subsection{\add{Comprehensive Candidate Pool for Validation}}

\add{To examine whether a richer representation would improve performance, we constructed a broader candidate pool of commonly used time-series descriptors for waveform shape and dynamics. Let $x = [x_1, x_2, \dots, x_T]$ denote a single-channel waveform segment consisting of $T$ samples, where $\mu$ and $\sigma$ represent the mean and standard deviation of the segment, respectively. Let $p \in \{1, 2, \dots, T\}$ be the index of the detected peak. The first- and second-order discrete derivatives of the signal are denoted as $\Delta x_i = x_{i} - x_{i-1}$ and $\Delta^2 x_i = \Delta x_{i} - \Delta x_{i-1}$.}

\add{Specifically, in addition to the 10 features we extracted above for each EOG channel, we extracted the following categories of features to form the complete candidate pool:}

\begin{itemize}
    \item \add{Morphological Descriptors: Segment length $T$, and the normalized peak position calculated as $p / (T - 1)$. }
    \item \add{Distributional and Amplitude Statistics: Mean $\mu$, standard deviation $\sigma$, median, minimum $\min(x)$, maximum $\max(x)$, peak-to-peak amplitude $\max(x) - \min(x)$, skewness $\frac{1}{T} \sum_{i=1}^{T} (\frac{x_i - \mu}{\sigma})^3$, and kurtosis $\frac{1}{T} \sum_{i=1}^{T} (\frac{x_i - \mu}{\sigma})^4$.}
    \item \add{Energy Descriptors: Total signal energy $E = \sum_{i=1}^{T} x_i^2$, and the absolute area under the curve $\text{AUC} = \sum_{i=1}^{T} |x_i|$. We also compute the left-half energy $E_{L} = \sum_{i=1}^{p} x_i^2$ and right-half energy $E_{R} = \sum_{i=p}^{T} x_i^2$, which define the energy asymmetry as $\frac{E_{R} - E_{L}}{E_{L} + E_{R}}$.}
    \item \add{Derivative Descriptors: To capture local temporal dynamics, we calculate the gradient mean $\mu_{\Delta x}$, gradient standard deviation $\sigma_{\Delta x}$, mean absolute gradient $\frac{1}{T-1} \sum_{i=2}^{T} |\Delta x_i|$, and maximum absolute gradient $\max(|\Delta x|)$. Similarly, for the second-order derivative, we extract the mean absolute value $\frac{1}{T-2} \sum_{i=3}^{T} |\Delta^2 x_i|$ and maximum absolute value $\max(|\Delta^2 x|)$. Finally, we compute the gradient zero-crossing rate (ZCR) of $\Delta x$.}
\end{itemize}

\add{Together with the 10 compact descriptors, this comprehensive candidate pool contains 32 features per channel, resulting in a 64-dimensional feature vector for two-channel EOG fusion.}

\subsection{\add{Feature Evaluation and Selection}}

\add{To finalize our feature representation, we evaluated our proposed 20-dimensional compact set against both its single-channel variants and the comprehensive 64-dimensional candidate pool. The single-channel baselines extract the 10 compact features from only the left or right EOG channel. In contrast, the expanded representation utilizes the full 32-feature candidate pool per channel, totaling 64 features.}

\add{Our evaluation shows that two-channel fusion improves vergence classification. The compact two-channel representation achieved an accuracy of 74.95 $\pm$ 7.38\%, outperforming the individual single-channel variants (left channel only: 69.03 $\pm$ 9.16\%; right channel only: 66.30 $\pm$ 9.43\%). Furthermore, expanding the feature space to the larger 64-dimensional candidate pool yielded only a modest performance gain, reaching 75.74 $\pm$ 7.17\%. Consequently, we adopted the 20-dimensional compact representation for our main pipeline. This design choice preserves most of the performance of the richer candidate pool while minimizing feature dimensionality and computational overhead.}

\begin{table}[t]
\centering
\small
\begin{tabular}{lccc}
\hline
\textbf{Feature Representation} & \textbf{Features per Channel} & \textbf{Total Dimensions} & \textbf{Accuracy} \\
\hline
Single-channel (Left), Compact & 10 & 10 & 69.03 $\pm$ 9.16\% \\
Single-channel (Right), Compact & 10 & 10 & 66.30 $\pm$ 9.43\% \\
Two-channel fusion, Compact & 10 & 20 & 74.95 $\pm$ 7.38\% \\
Two-channel fusion, Full candidate pool & 32 & 64 & 75.74 $\pm$ 7.17\% \\
\hline
\end{tabular}
\caption{\add{Feature selection results for six-gesture vergence classification.}}
\label{tab:feature_selection}
\end{table}
\section{\add{Architecture and Implementation Details of Baseline Models}}

\label{sec:appendix_baseline}

\add{The Random Forest baseline based on features uses the 20-dimensional feature vector described in Appendix~\ref{sec:appendix_feature}. The deep learning baselines use the same filtered two-channel EOG segments directly, where each segment is resampled to 160 time points, producing a $2 \times 160$ input tensor.}

\add{For within-session evaluation, models were evaluated using five-fold cross-validation within each user session, with folds grouped by collection round. For cross-session evaluation, models were trained in one session and tested in the other session after remounting of the glasses for the same user, with both transfers evaluated. For cross-user evaluation, each target user-session was held out for testing, and models were trained only on sessions from the remaining users. To avoid label leakage, hyperparameters were tuned using only the training data; the held-out test labels were used only for final evaluation.}

\add{The 1D-CNN baseline consists of two temporal convolution blocks followed by a six-class classifier. The first block applies a 1D convolution to the two-channel input, followed by batch normalization, ReLU activation, max pooling, and dropout. The second block applies another 1D convolution followed by batch normalization, ReLU activation, and global average pooling. The pooled representation is then passed to a fully connected layer for a six-class prediction.}

\add{To capture longer-term temporal dependencies, the LSTM baseline treats the same $2 \times 160$ EOG segment as a temporal sequence with two input features per time step. The model uses one or two LSTM layers, optionally bidirectional, and routes the final time-step output to a linear six-class classifier.}

\add{Both deep learning baselines were trained using the AdamW optimizer and cross-entropy loss. For each evaluation split, we executed 50 Optuna Tree-structured Parzen Estimator (TPE) trials to automatically tune the hyperparameters. As shown in Table~\ref{tab:baseline_model_details}, the shared search space included learning rate, weight decay, batch size, and dropout rates. CNN-specific hyperparameters included the number of convolution channels and kernel sizes, while LSTM-specific hyperparameters included hidden size, number of recurrent layers, and bidirectionality.}

\begin{table}[t]
\centering
\small
\begin{tabular}{lp{0.68\linewidth}}
\hline
\textbf{Hyperparameter} & \textbf{Search space} \\
\hline
Learning rate & Log-uniform from $10^{-4}$ to $3 \times 10^{-3}$. \\
Weight decay & Log-uniform from $10^{-6}$ to $10^{-2}$. \\
Batch size & $\{16, 32, 64\}$. \\
CNN channels & First convolution: $\{8, 16, 32, 64\}$; second convolution: $\{16, 32, 64, 96\}$. \\
CNN kernel sizes & First convolution: $\{5, 7, 9, 11\}$; second convolution: $\{3, 5, 7\}$. \\
CNN dropout & Uniform from 0 to 0.5. \\
LSTM hidden size & $\{16, 32, 64, 96\}$. \\
LSTM layers & $\{1, 2\}$. \\
LSTM bidirectionality & $\{\text{unidirectional}, \text{bidirectional}\}$. \\
LSTM dropout & Uniform from 0 to 0.5; applied only when more than one recurrent layer is used. \\
\hline
\end{tabular}
\caption{\add{Hyperparameter search spaces for the 1D-CNN and LSTM baseline models. Each model was tuned using 50 Optuna TPE trials per evaluation split, using training data only.}}
\label{tab:baseline_model_details}
\end{table}
\section{\add{Evaluation of the Artifact-Removal Pipeline: Performance and Calibration Burden}}
\label{sec:appendix_motion_artifacts}
\add{This section provides additional details for the motion artifact removal evaluation described in Sec.~\ref{sec:motion_eval}. We investigate the impact of varying training data volumes (vergence and artifact rounds) on system performance and assess the pipeline's cross-user generalizability.}

\subsection{\add{Evaluation protocol and metrics}}

\add{Consistent with the real-time scenarios, all evaluations employed a 2-second moving window. To preclude data leakage among highly correlated adjacent windows, we enforced training-test splits at the recording or round level. System performance is quantified using three metrics: overall accuracy, vergence false rejection rate (the proportion of true vergence windows classified as artifacts), and artifact false acceptance rate (the proportion of artifact/noise windows incorrectly identified as vergence).}

\add{We investigated two settings. The personalized setting trained the model from scratch exclusively on the target participant's data. The cross-user adaptation setting pre-trained the model on data from all other participants before fine-tuning it with varying amounts of target-user data. Together, these settings demonstrate the registration burden for a personalized model and the feasibility of adapting a generalized model to new users.}

\subsection{\add{Personalized artifact-removal performance}}

\begin{figure*}
\includegraphics[width=0.98\linewidth]{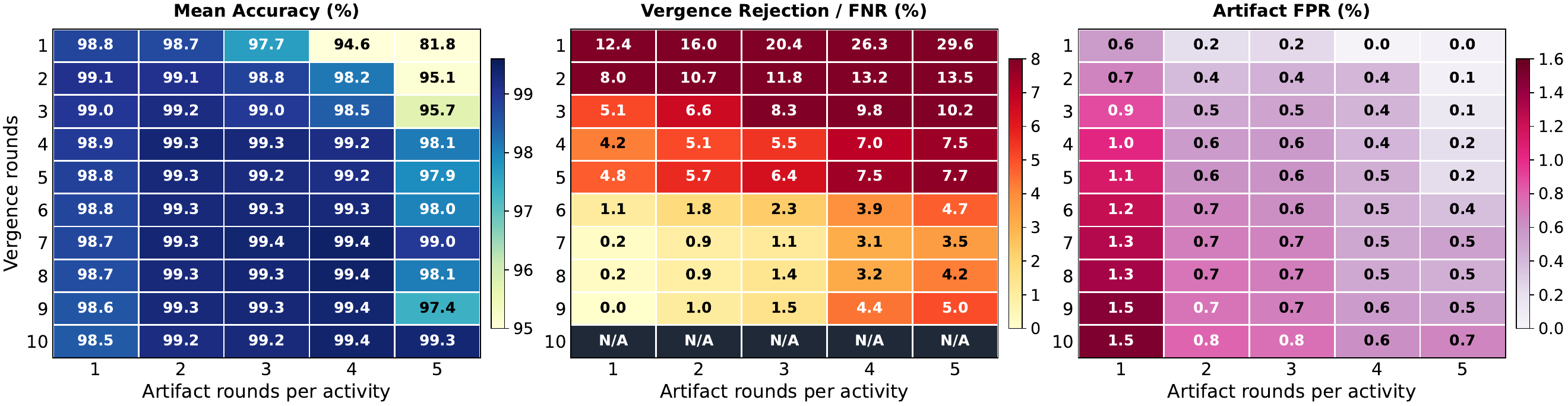}
    \caption{\add{\textbf{Training-data budget analysis for the personalized artifact-removal model.} Each cell reports the mean performance across all six participants for models trained from scratch using a specific number of vergence and artifact rounds.}}
    \label{fig:appendix_motion_per_user}
\end{figure*}

\add{Fig.~\ref{fig:appendix_motion_per_user} illustrates the personalized data-budget analysis. Each cell in the matrix represents a specific training-data budget, defined by the number of vergence rounds (horizontal axis) and artifact rounds (vertical axis). For each configuration, we report the accuracy, false negative rate, and false positive rate.}

\add{Our results reveal that increasing the volume of vergence training data primarily mitigates the false rejection of intended inputs. With a fixed budget of one artifact round, scaling the vergence training data from one to six rounds reduced the vergence false rejection rate from 12.36\% to 1.08\%. Concurrently, the artifact false acceptance rate remained stable and low, shifting only from 0.56\% to 1.23\%. This demonstrates that supplementary vergence examples enable the model to better preserve vergence inputs without compromising its ability to reject noise.}

\add{Guided by this analysis, we adopted six vergence rounds and one artifact round (V6/A1) as the primary operating point for training our main evaluation model. This configuration strikes a highly practical balance between system robustness and user burden, translating to the concise 6-minute total registration time for new users: 3 minutes for six rounds of vergence data, and an additional 3 minutes to collect one artifact calibration round for each of the six motion/noise activities. It yielded an overall accuracy of 98.77 $\pm$ 1.26\%, a false rejection rate of vergence of 1.08 $\pm$ 1.55\% and a false positive rate of 1.23 $\pm$ 1.26\%. While incorporating additional artifact rounds into the training set could further suppress false acceptances, it would linearly increase the data collection burden. Thus, the V6/A1 setting offers an optimal data collection protocol while maintaining minimal error rates.}

\subsection{\add{Cross-user generalization and target-user adaptation}}

\begin{figure*}
\includegraphics[width=0.98\linewidth]{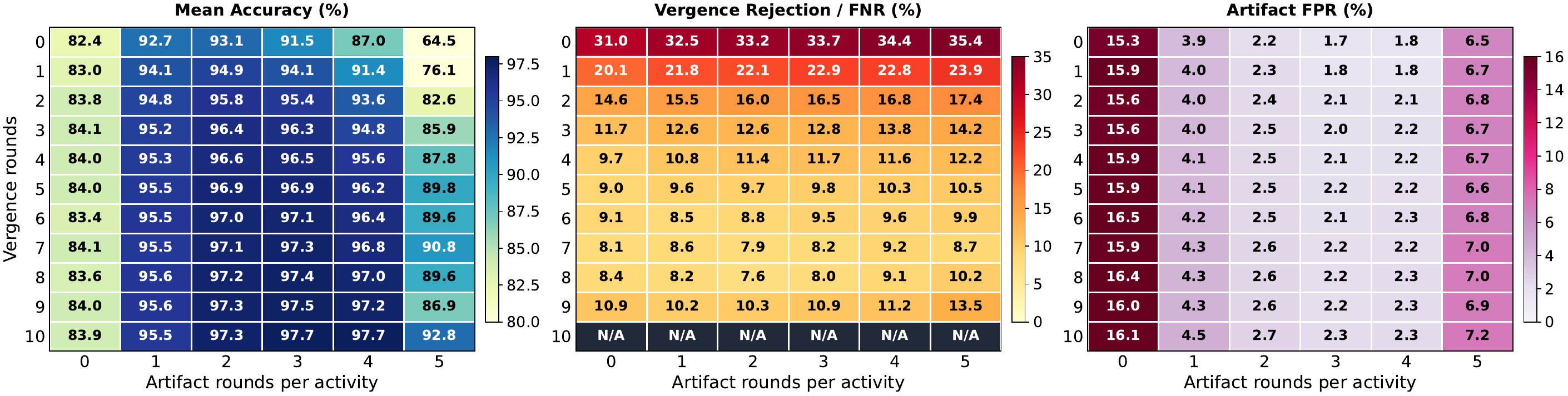}
    \caption{\add{\textbf{Performance of the cross-user artifact-removal model with target-user adaptation.} The model is pre-trained on data from other participants and subsequently adapted using varying amounts of target-user data. Each cell reports the mean performance across all six participants for cross-user models adapted with a specific number of target-user vergence and artifact rounds.}}
    \label{fig:appendix_motion_cross_user}
\end{figure*}

\add{Fig.~\ref{fig:appendix_motion_cross_user} details the performance of the cross-user adaptation setting. Here, the model was pre-trained on the remaining participants and subsequently adapted using variable amounts of data from the target user. In a zero-shot scenario, the model achieved only 82.41\% accuracy, severely affected by a 31.02\% vergence false rejection rate and a 15.33\% artifact false positive rate. This performance drop highlights the inter-user variance in motion and artifact patterns, likely stemming from individual differences in sensor contact, facial anatomy, and motion execution.}

\add{Introducing target-user vergence data alone mitigated false rejections but failed to adequately suppress artifact acceptances. For example, incorporating six vergence rounds for the target-user (without artifact data) reduced the false rejection rate of vergence to 9.10\%, yet the false positive rate remained high at 16.51\%. This indicates that while positive vergence examples help anchor the intended signal, they lack the negative constraints necessary to accurately model user-specific noise profiles.}

\add{In contrast, injecting even a single artifact round from the target participant into the adaptation process enhanced noise rejection. By combining six vergence rounds with one artifact round from the target user, the adapted cross-user model achieved 95.55\% accuracy, with an 8.53\% vergence false rejection rate and a 4.16\% artifact false positive rate. This result shows the importance of sampling at least a minimal amount of target-user artifact data for robust noise suppression.}

\add{In summary, these evaluations justify the adoption of the personalized V6/A1 setting, where the model was trained from scratch using six rounds of vergence data and one round of artifact data per activity. Fully generalized cross-user models prove insufficiently reliable for real-world deployment, whereas a brief, 6-minute personalized data collection session to train the model from scratch yields reliable accuracy and minimal error across both metrics. The data-budget ablation confirms the V6/A1 configuration as a practical operating point that maximizes algorithmic reliability while respecting strict real-world usability constraints.}

\end{document}